\documentclass[%
 reprint,
superscriptaddress,
 amsmath,amssymb,
 aps,
]{revtex4-2}
\usepackage{graphicx}
\usepackage{dcolumn}
\usepackage{bm}

\usepackage{lipsum}
\usepackage{xcolor}
\usepackage[export]{adjustbox}
\begin{document}

\preprint{APS/123-QED}

\title{Chaotic dynamics under the influence of synthetic magnetic field\\in optomechanical system}
\author{Souvik Mondal}
 \email{souvikjuetce95@kgpian.iitkgp.ac.in}
 \affiliation{%
Electronics and Electrical Communication Engineering Department, IIT Kharagpur, West Bengal, 721302, India}%
 \author{Murilo S. Baptista}%
 \affiliation{%
School of Natural and Computing Sciences, University of Aberdeen, Aberdeen AB24 3UE, UK}%
 \author{Kapil Debnath}%
 \email{kapil.debnath@abdn.ac.uk}
\affiliation{%
Electronics and Electrical Communication Engineering Department, IIT Kharagpur, West Bengal, 721302, India}%
 \affiliation{%
School of Natural and Computing Sciences, University of Aberdeen, Aberdeen AB24 3UE, UK}%
\begin{abstract}
The optomechanical systems produce chaotic behaviour due to nonlinear interaction between photons and phonons, and the same systems are used to understand the synthetic fields as well. Here, we report on the study of chaotic behaviour in the presence of a phononic synthetic magnetic field in a closed loop configuration consisting of a single optical mode and two mechanical modes. The modulation phase of the mechanical coupling between the two mechanical modes plays a critical role in determining the mechanical and optical intensity dynamics in the nonlinear regime. Our study shows the dark mode breaking effect in the presence of a synthetic magnetic field, which brings about a complex way of mechanical energy exchange that causes the cavity field to alternate between chaotic and regular behaviour periodically in temporal domain. However in the stronger nonlinear regime the temporal dynamics demonstrate predominantly chaotic behaviour. Besides, with the advent of advanced fabrication technologies, this study holds promises in developing phase tunable integrated low-power chaotic light sources to support efficient optical secure communication systems. 
\end{abstract}
\maketitle
\section{\label{sec:level1}Introduction}
Cavity optomechanical systems allow significant interaction between the light field and mechanical vibrations which are being utilized to show various rich classical and quantum phenomena \cite{aspelmeyer2014cavity}. Since the nature of the interaction is fundamentally nonlinear \cite{hossein2006characterization,carmon2005temporal,mondal2023controllable}, the optomechanical cavity provides an ideal platform to observe chaotic phenomenon \cite{carmon2007chaotic,navarro2017nonlinear}. Chaos in the optomechanical cavity occurs in the highly nonlinear regime when the driving power of the cavity becomes sufficiently high \cite{bakemeier2015route}. The dynamics in the chaotic regime has a noise-like behaviour originating from a purely deterministic set of equations and is extremely sensitive to infinitesimal changes in the initial conditions \cite{thompson1990nonlinear}.

Many studies have been conducted regarding a deeper understanding of the chaotic behaviour in different configurations of optomechanical systems and in controlling the chaotic behaviour in those systems. For example, controllability in chaotic motions in an optomechanical system can be achieved through the pump and probe fields applied to the system \cite{ma2014formation,monifi2016optomechanically}. In a multimode optomechanical configuration, the tunability of the chaotic motions are shown in an optical Parity-Time ($\mathcal{PT}$) symmetry optomechanical system \cite{lu2015p}, anti-$\mathcal{PT}$ symmetric optomechanical structure \cite{huang2021tunable} and in a passive double cavity optomechanical configuration \cite{bai2023tunable,zhang2024loss}. Zhang \textit{et al.} \cite{zhang2020intermittent,zhang2021nonreciprocal}, in his studies, showed that the intermittent chaos exists in an optomechanical cavity and the non-reciprocal nature of chaotic motion is reflected in a spinning optomechanical cavity. Therefore, the platform of optomechanical systems promises to provide an integrated low-power tunable chaos-based communication system \cite{vanwiggeren1998communication,cuomo1993circuit,ren2013wireless}, random number generators\cite{uchida2008fast}, among other applications of chaos.

In a different context, optomechanical systems also provide a platform to understand synthetic (artificial) fields\cite{walter2016classical,mathew2020synthetic,chen2021synthetic,zapletal2019dynamically} that help to understand the topological phases of matter \cite{peano2015topological}, different exotic transport phenomena \cite{yang2015topological,fleury2016floquet} and others. Our considered system has phase-dependent mechanical coupling between two mechanical modes, which give rise to the phononic synthetic magnetic field and the two mechanical modes are coupled to a common optical mode \cite{schmidt2015optomechanical}. This type of closed-loop three-mode optomechanical configuration is utilized to show tunable optomechanically induced transparency (OMIT) \cite{lai2020tunable}, noise-tolerant entanglement \cite{lai2022noise}, and controllable generation of mechanical squeezing by breaking the dark mode effect \cite{PhysRevA.108.013516}; and also used to provide enhanced mass sensing \cite{tchounda2023sensor}. But chaos-based study is still unexplored in this particular optomechanical configuration. Thereby, our study aims to understand the nature of the chaotic dynamics and explore the controllable aspects of the chaotic dynamics in this configuration. The initial part of our study examines the mechanical dynamics for different mechanical coupling phases at a fixed driving power level and then we explain the dynamics in terms of nonlinear way of mechanical energy exchange process between the two hybridized mechanical modes. We studied the effect of such mechanical dynamics on the evolution of the optical intensity inside the cavity where the periodic appearances of chaos and regular behaviour in the temporal domain is observed, and the periodicity turned out to be governed by the mechanical coupling rate. The difference between chaotic and regular behavior is clarified by examining the nature of trajectories in phase space, the rate of separation of two infinitesimal close trajectories in phase space, and the optical and mechanical spectra. We also showed how the periodic chaos-regular dynamics are suppressed while the chaotic behaviour dominates in the higher driving power levels and demonstrated the mechanical coupling phase-dependent nature of the dynamics through a bifurcation plot. Therefore, the studies provided an idea of unique dynamics in this system at different power levels and the tunable nature of chaotic motions based on varying mechanical coupling phases. The system of study can be experimentally realized based on one-dimensional (1D) optomechanical crystal cavities \cite{fang2017generalized,lai2022noise} or circuit electro-mechanical system \cite{massel2011microwave,massel2012multimode,lai2022noise,lai2020tunable} and therefore, kept the values of the parameters in our study similar to \cite{fang2017generalized}.

The rest of the paper is organized as follows. In Section II, we mathematically model our optomechanical system and mention the procedure to quantify chaotic behaviour. The mechanical dynamics for various situations are being studied in Sec III. In Sec IV the corresponding behaviour of light intensity inside the cavity is studied for the case of low to high driving power levels. Finally, we summarize our results in Sec V.
\section{MATHEMATICAL MODELLING}
\begin{figure}
    \centering
    \includegraphics[width=0.55\linewidth]{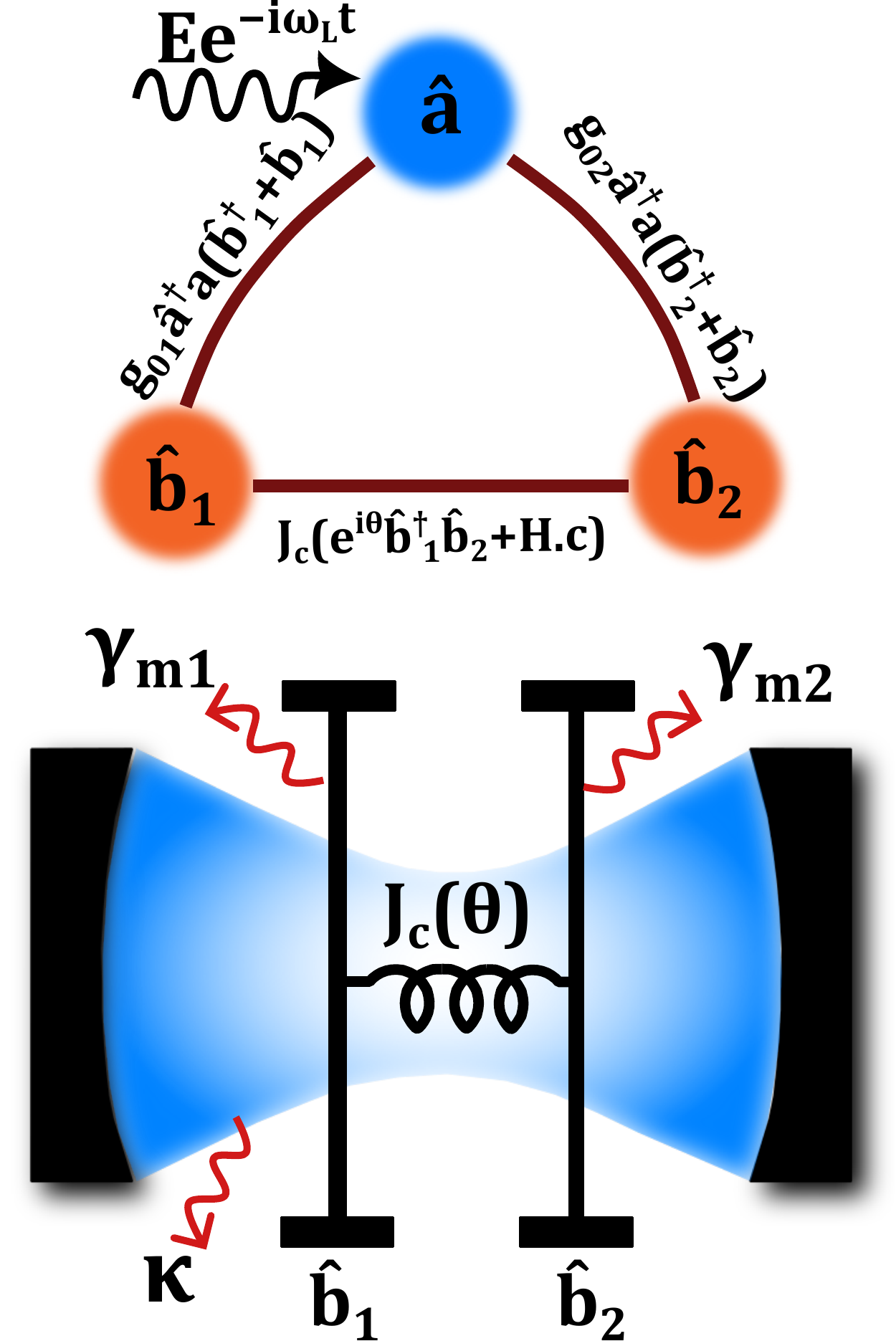}
    \caption{The schematic of our configuration consisting of single optical mode $\hat{a}$ coupled to the mechanical modes $\hat{b}_1$ and $\hat{b}_2$ with resonance frequency $\omega_{m_1}$ and $\omega_{m_2}$ respectively. The mechanical modes are also being coupled mechanically of strength $J_c$ with phase modulation $\theta$.}
    \label{fig1}
\end{figure}

The configuration of our system is shown in Fig. \ref{fig1} where there is a single optical mode coupled to two mechanical modes. The mechanical modes are weakly coupled with each other, with the coupling being phase-modulated \cite{lai2020tunable,lai2022noise,PhysRevA.108.013516,tchounda2023sensor}. Therefore, the configuration forms a closed loop structure as in Fig. \ref{fig1}, and the phase modulation of the mechanical coupling creates a synthetic magnetic flux in the system. Assuming $\hbar=1$ the Hamiltonian of the system in the rotating frame of driving laser with frequency $\omega_L$ is written as
\begin{subequations}
   \begin{eqnarray}
    H_{\text{free}}&=& -\Delta \hat{a}^{\dagger}\hat{a}+\sum_{j=1,2} \omega_{m_j}\hat{b}^{\dagger}\hat{b},\\
    H_{\text{int}}&=&J_c(e^{i\theta}\hat{b}^{\dagger}_1\hat{b}_2+e^{-i\theta}\hat{b}_1\hat{b}^{\dagger}_2)\nonumber\\&&+\sum_{j=1,2} g_{0_j}\hat{a}^{\dagger}\hat{a}(\hat{b}^{\dagger}_j+\hat{b}_j),\\
    H_{\text{drive}}&=& iE(\hat{a}^{\dagger}-\hat{a}),\\
    H_{\text{total}}&=&H_{\text{free}}+H_{\text{int}}+H_{\text{drive}}.
\end{eqnarray} 
 \label{eq1}
\end{subequations}

The free Hamiltonian $H_{\text{free}}$ of the system includes the optical modes (represented by the annihilation and destruction operator $\hat{a}^\dagger$ and $\hat{a}$ respectively) with detuning $\Delta$ and two mechanical modes (represented by the operators $\hat{b}^\dagger_{1,2}$ and $\hat{b}_{1,2}$ respectively) with resonance frequency $\omega_{m_{1,2}}$. The detuning is defined by the deviation of the laser frequency $\omega_L$ from the optical cavity resonance $\omega_{cav}$ such that $\Delta=\omega_L-\omega_{cav}$. The interaction terms in $H_\text{int}$ consist of optomechanical interaction term $g_{0_{1,2}}$  of the single optical mode with the two mechanical modes and the mechanical coupling term $J_c$ ($\ll\omega_m$) with phase modulation $\theta$ between the two mechanical modes. Lastly, the $H_{\text{drive}}$ contains the laser driving strength $E$. In terms of input power $P_{in}$ the drive amplitude is written as $E=\sqrt{\frac{2\kappa P_{in}}{\hbar\omega_L}}$ where $\kappa$ is the decay rate of the optical cavity. The semi-classical Heisenberg-Langevin dynamical equations for the complex light field ($a\equiv\langle\hat{a}\rangle$)  and the dimensionless complex mechanical displacement ($b_{1,2}\equiv\langle\hat{b}_{1,2}\rangle$) without consideration of any classical noises are given by
\begin{subequations}
\begin{eqnarray}
\dot a&=&i\tilde{\Delta}a-\frac{\kappa}{2}a+E,\\
\dot b_1&=&-\left(i\omega_{m_1}+\frac{\gamma_{m_1}}{2}\right)b_1+iJ_ce^{i\theta}b_2+ig_{0_1}|a|^2,\\
\dot b_2&=&-\left(i\omega_{m_2}+\frac{\gamma_{m_2}}{2}\right)b_2+iJ_ce^{-i\theta}b_1+ig_{0_2}|a|^2.
\end{eqnarray}
\label{eq2}
\end{subequations}
The effective detuning due to the optomechanical interaction is given by $\tilde{\Delta}=\Delta+2\left[\sum_{j=1,2}g_{0_j}\text{Re}(b_j)\right]$ and the decay rate of the mechanical vibration in each mechanical resonator is denoted by $\gamma_{m_{1,2}}$. Under steady-state condition, the average values of the variables are written as
\begin{subequations}
    \begin{eqnarray}
    \bar a&=&\frac{E}{-i\bar{\tilde{\Delta}}+\frac{\kappa}{2}},\\
    \bar b_1&=&\frac{iJ_ce^{i\theta}}{i\omega_{m_1}+\frac{\gamma_{m_1}}{2}}\bar b_2+\frac{ig_{0_1}|\bar a|^2}{i\omega_{m_1}+\frac{\gamma_{m_1}}{2}},\\
    \bar b_2&=&\frac{iJ_ce^{-i\theta}}{i\omega_{m_2}+\frac{\gamma_{m_2}}{2}}\bar b_1+\frac{ig_{0_2}|\bar a|^2}{i\omega_{m_2}+\frac{\gamma_{m_2}}{2}},
\end{eqnarray}
\label{eq2.1}
\end{subequations}
where $\bar{\tilde{\Delta}}$ is the effective detuning under steady-state condition given by $\bar{\tilde{\Delta}}=\Delta+2\left[\sum_{j=1,2}g_{0_j}\text{Re}(\bar b_j)\right]$. The Eq. (\ref{eq2}) has been solved numerically since chaos appears in the highly nonlinear regime and analytic solutions in the nonlinear regime are generally difficult to obtain.

The characterization of the chaotic behaviour would require the knowledge of the dynamics of the perturbed quantities $\vec{\delta}=(\delta a_r,\delta a_{im},\delta b_{1_r},\delta b_{1_{im}},\delta b_{2_r},\delta b_{2_{im}})$ , which is given by the variational equation $\dot{\vec{\delta}}=M\vec{\delta}$ after linearizing Eq. (\ref{eq2}). The matrix $M_{6\times6}$ is denoted by
\begin{widetext}
    \begin{equation}
M_{6\times6}=
\begin{pmatrix}
    -\frac{\kappa}{2} & -\tilde{\Delta} & -2g_{0_1}\text{Im}(a) & 0 & -2g_{0_2}\text{Im}(a) & 0\\
    \tilde{\Delta} & -\frac{\kappa}{2} & 2g_{0_1}\text{Re}(a) & 0 & 2g_{0_2}\text{Re}(a) & 0\\
    0 & 0 & -\frac{\gamma_{m_1}}{2} & \omega_{m_1} & -J_c\text{sin}\theta & -J_c\text{cos}\theta \\
    2g_{0_1}\text{Re}(a) & 2g_{0_1}\text{Im}(a) & -\omega_{m_1} & -\frac{\gamma_{m_1}}{2} & J_c\text{cos}\theta & -J_c \text{sin}\theta \\
    0 & 0 & J_c\text{sin}\theta & -J_c\text{cos}\theta & -\frac{\gamma_{m_2}}{2} & \omega_{m_2} \\
    2g_{0_2}\text{Re}(a) & 2g_{0_2}\text{Im}(a) & J_c\text{cos}\theta & J_c\text{sin}\theta & -\omega_{m_2} & -\frac{\gamma_{m_2}}{2} \\ 
\end{pmatrix},
\label{eq2.2}
\end{equation}
\end{widetext}
and the subscript `r' and `im' denote the real and imaginary part of the perturbation of the semi-classical variables $a$, $b_1$ and $b_2$, respectively. The chaotic dynamics are verified by computing the largest Lyapunov exponent (LLE), which denotes the rate of separation of the infinitesimally close trajectories \cite{eckmann1985ergodic,araujo2019lyapunov} along the most unstable direction. We are interested in the perturbed value of the optical intensity given by $\delta I=[\text{Re}(a)+\delta a_r]^2+[\text{Im}(a)+\delta a_{im}]^2-|a|^2$ and therefore for a time duration $t=0$ to $t=T$, the LLE is defined by
\begin{equation}
    \text{LLE}=\lim_{T\to 0}\frac{1}{T}\text{ln} \bigg| \frac{\delta I(T)}{\delta I(0)}\bigg|,
\end{equation}
where $\delta I(T)$ is obtained by solving  Eq. (\ref{eq2}) and the variational equation $\dot{\vec{\delta}}=M\vec{\delta}$ with $\delta I(0)\to 0$.  But for a longer time duration, the LLE is rewritten as
\begin{flalign}
\text{LLE}&=\lim_{N\to\infty}\frac{1}{N}\lim_{\tau\to 0}\frac{1}{\tau}\text{ln} \bigg| \frac{\delta I(N\tau)}{\delta I((N-1)\tau)}\frac{\delta I((N-1)\tau)}{\delta I((N-2)\tau)}\nonumber\\
&
\cdot\cdot\cdot\frac{\delta I(\tau)}{\delta I(0)}\bigg|\nonumber\\
&=\lim_{N\to\infty}\frac{1}{N}\sum_{k=1}^{k=N}\lim_{\tau\to 0}\frac{1}{\tau}\text{ln} \bigg| \frac{\delta I(k\tau)}{\delta I((k-1)\tau)}\bigg|,
\end{flalign}
where $N$ is the total number of iterations involved in solving Eq. (\ref{eq2})
and the variational equation, with each integration time step given by $\tau$ ($\tau$ is chosen to be $1/100^{\text{th}}$ times of the time period of the mechanical oscillation ) such that the total time duration $T=N\tau$. Now, after each integration time step the perturbed quantities are normalized by the norm, that is, $\vec{\delta}(k\tau)=\vec{\delta}(k\tau)/\|\vec{\delta}(k\tau)\|$ , such that the perturbed trajectory remains close to the non-perturbed trajectory. A positive value of LLE indicates chaotic oscillation, while the negative implies regular motions.

In our study, the values of the parameters are based on the experimental work \cite{fang2017generalized} where two optomechanical cavities in a 1D optomechanical crystal are coupled both optically and mechanically. The closed loop configuration of Fig. \ref{fig1} can be obtained by adiabatically eliminating one of the optical modes \cite{fang2017generalized,lai2022noise} in the large detuned regime. Therefore, the values of the parameters are chosen as $\omega_{m_1}=\omega_{m_2}=\omega_{m}=2\pi\times6$ GHz, $\kappa=2\pi\times1.03$ GHz, $g_{0_1}=g_{0_2}=g_0=2\pi\times0.76$ MHz and $\gamma_{m_1}=\gamma_{m_2}=\gamma_m=2\pi\times1$ MHz. The mechanical coupling rate is fixed at $J_c=20.6\;\text{MHz} (\ll\omega_m)$ and the driving laser is operated in the blue detuned regime, which is set at $\Delta=\omega_m$.
\section{Temporal mechanical dynamics}
In this section, we discussed the dark mode breaking effect through generation of synthetic magnetic field in the linear regime. In addition, we study in detail the nature of the mechanical dynamics of the two mechanical resonators for different mechanical coupling phases.
\subsection{Linearized Hamiltoninan}
Here, the Hamiltoninan in Eq. (\ref{eq1}) is being linearized by considering the operators as fluctuation operators around steady-state values, which is $\hat{o}=\bar o+\delta\hat{o}$ (where $\hat{o}=\hat{a},\hat{b_1},\hat{b_2}$) and $\bar o$ is given in Eq. (\ref{eq2.1}). After linearization, the rotating wave approximation (RWA) is applied under the situation of blue-detuned driving of the cavity to obtain the resultant Hamiltonian as 
\begin{eqnarray}
    H_{\text{RWA}}&=&-\Delta\delta\hat{a}^{\dagger}\delta\hat{a}+\sum_{j=1,2}\omega_{m_j}\delta\hat{b_j}^{\dagger}\delta\hat{b_j}+J_c(e^{i\theta}\delta\hat{b_1}^{\dagger}\delta\hat{b_2}\nonumber\\
    &&+e^{-i\theta}\delta\hat{b_1}\delta\hat{b_2}^{\dagger})+\sum_{j=1,2}G_j(\delta\hat{b_j}^{\dagger}\delta\hat{a}^{\dagger}+\delta\hat{b_j}\delta\hat{a}),
    \label{eq4}
\end{eqnarray}
where $G_j$ is the effective optomechanical strength defined by $G_{1,2}=g_{0_{1,2}}\bar a$ and $\bar a$ assumed to be real. In the absence of synthetic magnetism ($J_c=0$) the mechanical modes can be represented with two hybridized modes, which are the bright ($\hat{b}_+$) and the dark mode ($\hat{b}_-$) and they are defined by
\begin{subequations}
    \begin{eqnarray}
    \hat{b}_+=\frac{G_1\delta\hat{b}_1+G_2\delta\hat{b}_2}{\sqrt{G_1^2+G_2^2}},\\
    \hat{b}_-=\frac{G_2\delta\hat{b}_1-G_1\delta\hat{b}_2}{\sqrt{G_1^2+G_2^2}},
\end{eqnarray}
\end{subequations}
where the operators satisfy the commutation relation $[\hat{b}_\pm,\hat{b}_\pm^{\dagger}]=1$. The bright mechanical mode couples with the cavity field with the effective optomechanical strength given by $G_+=\sqrt{G_1^2+G_2^2}$ whereas the dark mechanical mode does not interact with the cavity field. But in the presence of synthetic magnetism ($J_c\neq0$), the dark mode effect is broken \cite{lai2022noise,lai2020tunable,lai2020nonreciprocal} and therefore, the cavity field interacts with the dark mode as well. This can be understood by defining another two new modified mechanical mode operators, which are given by
\begin{subequations}
    \begin{eqnarray}
    \hat{\tilde{b}}_+&=&f_1\delta\hat{b}_1-f_2e^{i\theta}\delta\hat{b}_2,\\
    \hat{\tilde{b}}_-&=&f_2e^{-i\theta}\delta\hat{b}_1+f_1\delta\hat{b}_2,
\end{eqnarray}
\end{subequations}
where the operators follow $[\hat{\tilde{b}}_\pm,\hat{\tilde{b}}_\pm^{\dagger}]=1$ and $f_{1,2}$ is given by
\begin{flalign}
f_1=\frac{|\delta\tilde{\omega}|}{\sqrt{\delta\tilde{\omega}^2+J_c^2}},\;f_2=\frac{J_cf_1}{\delta\tilde{\omega}},
\end{flalign}
with $\delta\tilde{\omega}=\tilde{\omega}_--\omega_{m_1}$ and the resonance frequencies corresponding to the modified hybridized modes $\hat{\tilde{b}}_\pm$ are  defined by
\begin{eqnarray}
    \tilde{\omega}_\pm=\frac{\omega_{m_1}+\omega_{m_2}}{2}\pm\frac{\sqrt{(\omega_{m_1}-\omega_{m_2})^2+4J_c^2}}{2}.
\end{eqnarray}
In this scenario, the Hamiltonian in Eq. (\ref{eq4}) is written as
\begin{eqnarray}
H_{\text{RWA}}^\prime&=&-\Delta\delta\hat{a}^{\dagger}\hat{a}+\tilde{\omega}_+\hat{\tilde b}_+^{\dagger}\hat{\tilde b}_++\tilde{\omega}_-\hat{\tilde b}_-^{\dagger}\hat{\tilde b}_-+\tilde{G}_+(\hat{\tilde b}_+^{\dagger}\delta\hat{a}^{\dagger}+\nonumber\\&&
\hat{\tilde b}_+\delta\hat{a})+\tilde{G}_-(\hat{\tilde b}_-^{\dagger}\delta\hat{a}^{\dagger}+\hat{\tilde b}_-\delta\hat{a}),
\label{eq9}
\end{eqnarray}
where $\tilde G_+=f_1G_1- e^{- i\theta}f_2G_2$ and $\tilde G_-=f_1G_2+ e^{+ i\theta}f_2G_1$. As per the chosen values of the parameters in the previous section, $G_1=G_2=G$ and $\omega_{m_1}=\omega_{m_2}=\omega_m$ and thereby $\tilde G_\pm=G(1\pm e^{\mp i\theta})/\sqrt 2$. For $\theta=n\pi$ (where $n=0,1,2,...$) either of $\tilde G_+$ or $\tilde G_-$ is 0 which implies one of the hybridized mechanical modes $\hat{\tilde{b}}_\pm$ is coupled to the cavity field. Interestingly, for $\theta\neq n\pi$, the cavity field couples to both mechanical modes. The application of RWA and the linearization technique is being done to demonstrate the dark mode breaking effect and to explain some of the properties of the mechanical dynamics. Note that the mechanical dynamics, which are studied in the next section, are obtained from the full dynamical equation without using any approximation or linearization technique.
\subsection{Numerical results}
We solved Eq. (\ref{eq2}) numerically using Runge-Kutta method to obtain the solutions of the mechanical displacements of the oscillators. The initial condition of all dynamical variables was kept to zero. We kept the driving power fixed at $E=4000\omega_m$ in the numerical simulation and explored the behaviour of the dynamics for different phases $\theta$ of mechanical coupling $J_c$. The total observation period for the dynamics in all scenarios ranged from $1/\kappa \ll t < 1/\gamma_m$.
\begin{figure}
    \centering
    \includegraphics[width=\linewidth]{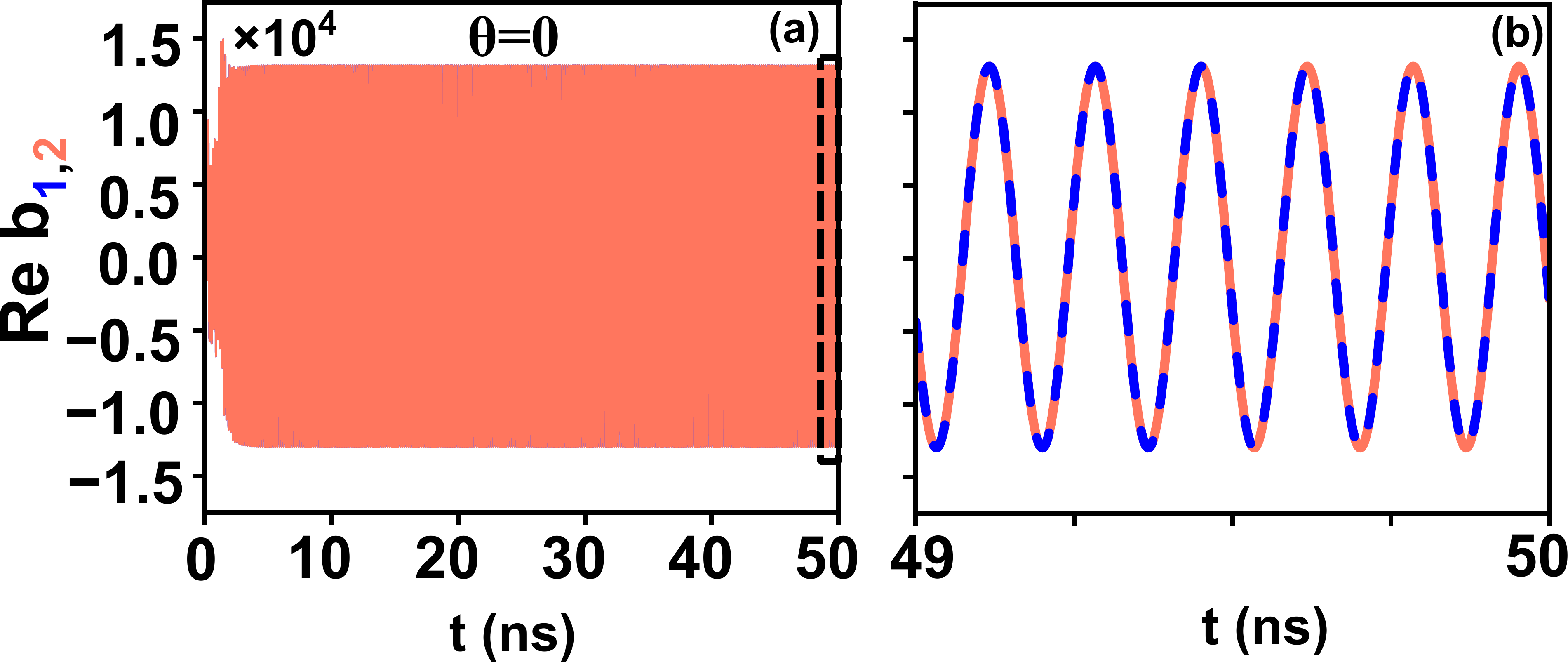}
    \caption{ (a) The mechanical dynamics $\text{Re} \;b_{1,2}$ of the two resonators for the mechanical coupling  phase of  $\theta=0$. (b) The magnified plot corresponding to the dashed box shows an in-phase oscillation with steady amplitude.}
    \label{fig2}
\end{figure}
\begin{figure}
    \centering
    \includegraphics[width=\linewidth]{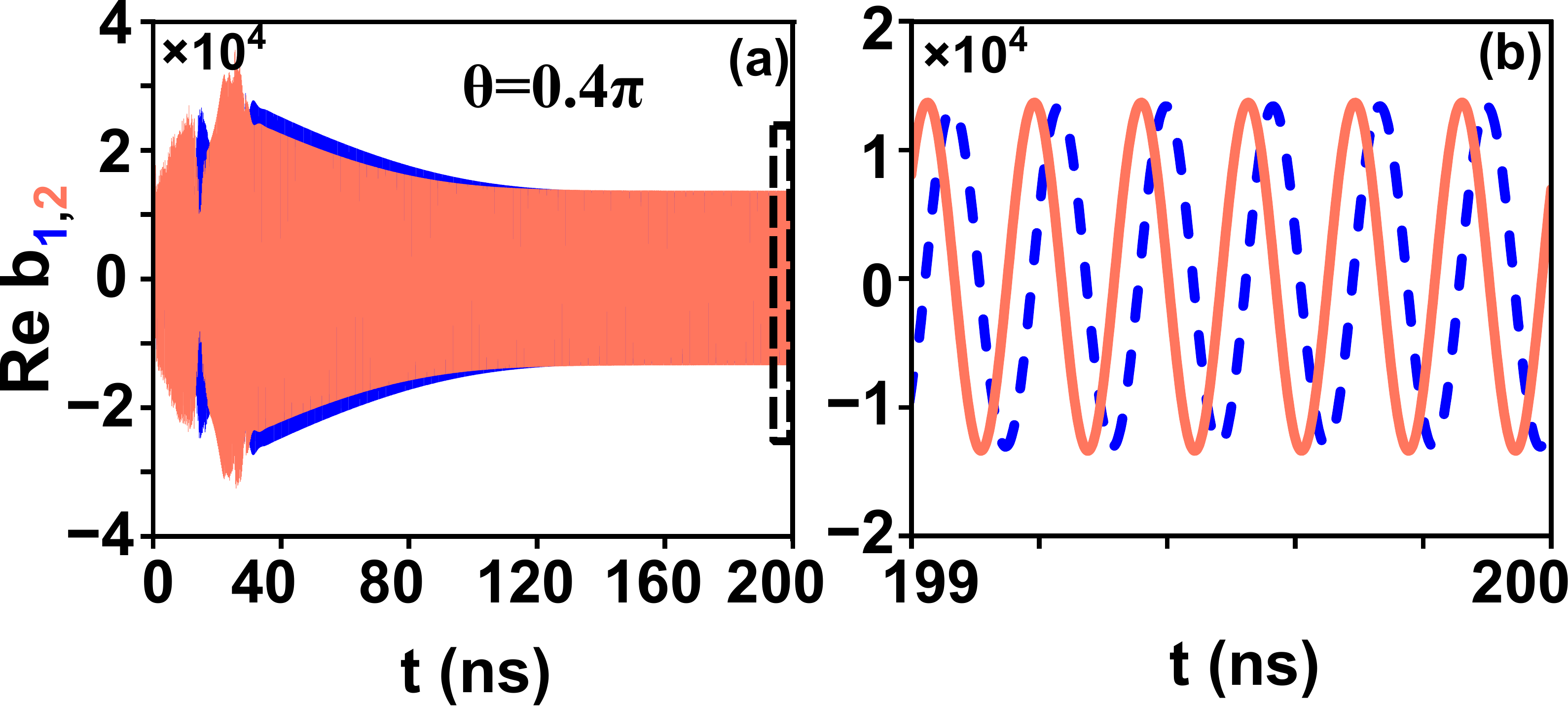}
    \caption{(a)The mechanical dynamics $\text{Re} \;b_{1,2}$ of the two resonators for the phase $\theta=0.4\pi$. (b) The magnified plot corresponding to the dashed box shows the oscillations of the two resonators but with a constant phase difference.}
    \label{fig3}
\end{figure}

In the first scenario, the mechanical dynamics $x_{1,2}\propto\text{Re}\;b_{1,2}$ is provided in Fig. \ref{fig2}(a) with the phase value of the mechanical coupling fixed at $\theta=0$, where the dynamics quickly settles into a steady oscillation with fixed amplitude. A zoomed-in view of a brief time segment from Fig. \ref{fig2}(a) is presented in Fig. \ref{fig2}(b), where the oscillations of the two mechanical modes are in-phase. This in-phase nature of the oscillations is attributed to the coherent nature of the mechanical coupling (purely real $J_c$). Now, as shown in Fig. \ref{fig3}(a), changing the mechanical coupling phase to $\theta = 0.4\pi$ (which also implies the introduction of imaginary mechanical coupling) maintains the steady dynamics similar to those in Fig. \ref{fig2}, with the exception of a constant phase difference, which is plotted in Fig. \ref{fig3}(b). But here, the transient dynamics is vastly different from Fig. \ref{fig2} and, we will see in the next scenario that this transient behaviour will become stable.
\begin{figure}
    \centering
    \includegraphics[width=\linewidth]{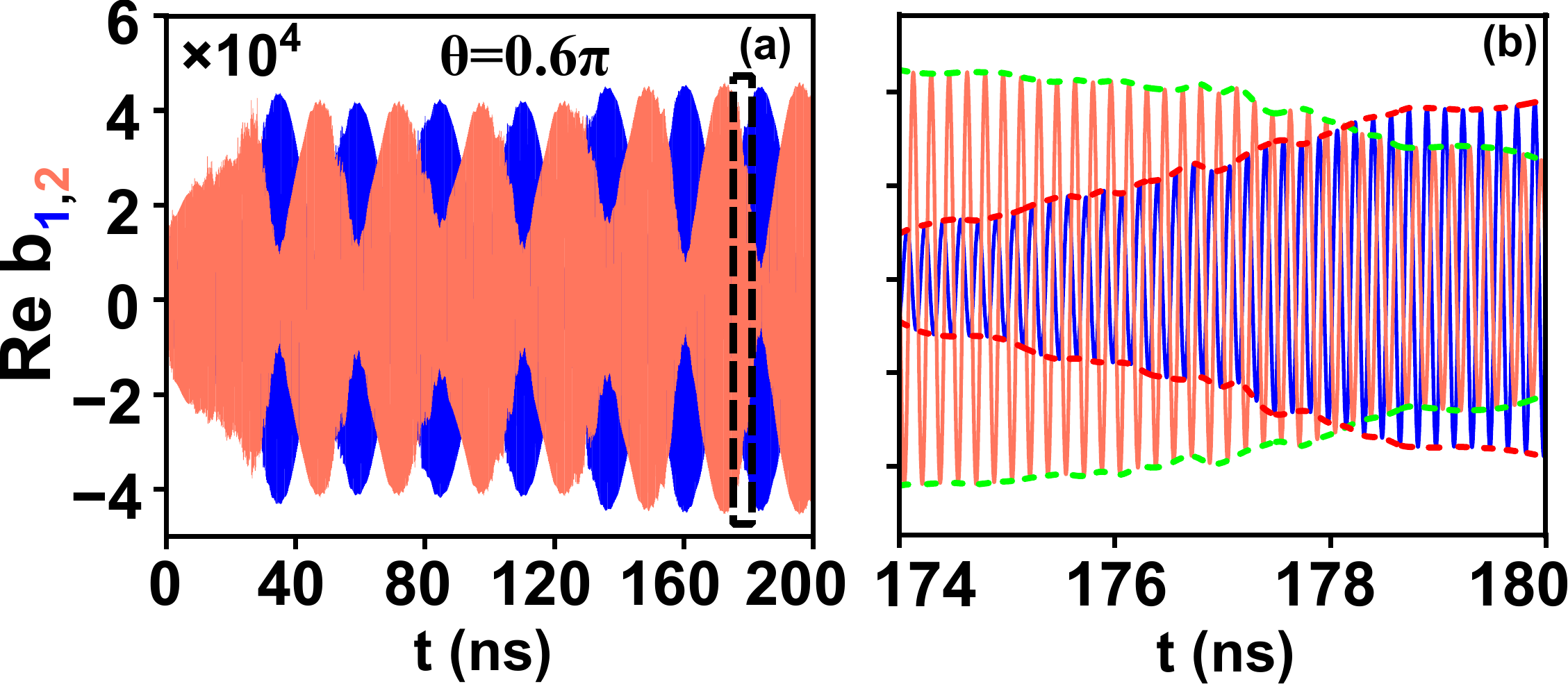}
    \caption{(a)The mechanical dynamics $\text{Re} \;b_{1,2}$ of the two resonators for the phase $\theta=0.6\pi$. (b) The magnified plot of the dashed box with the envelope of the oscillations shown by the red and green dashed line.}
    \label{fig4}
\end{figure}

In the second scenario, where the phase is tuned to $\theta=0.6\pi$, the dynamics in Fig. \ref{fig4}(a)  become far more interesting as a certain form of modulations appears in the slowly varying amplitude of mechanical oscillations. Following the envelope of the oscillations in Fig. \ref{fig4}(a), we see that it rises and dips in an irregular fashion in every cycle. Moreover, the zoomed-in view of the oscillations in Fig. \ref{fig4}(b), taken over a short time interval from Fig. \ref{fig4}(a), reveals that the envelope line exhibits non-uniform and abrupt characteristics. Overall the mechanical dynamics in Fig. \ref{fig4} indicates a strong optomechanical non-linearity.
\begin{figure}
    \centering
    \includegraphics[width=0.7\linewidth]{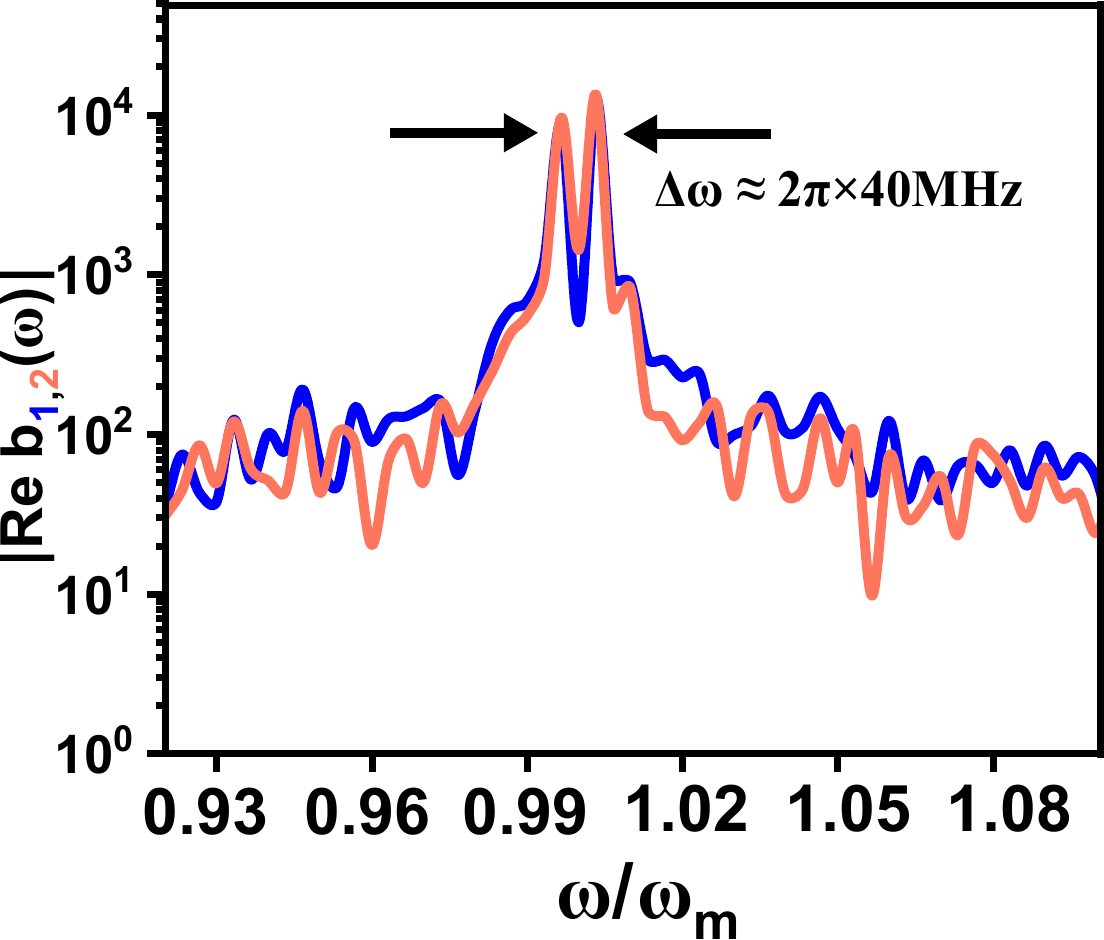}
    \caption{The mechanical spectrum of the resonators displaying two prominent peaks with a separation $\Delta\omega\approx2\pi\times40\text{MHz}$ and the existence of smaller peaks in a continuous manner.}
    \label{fig5}
\end{figure}

To understand more of the dynamics in Fig. \ref{fig4}, we numerically computed the displacement spectrum of the mechanical resonators in Fig. \ref{fig5} by performing a Fast Fourier Transform (FFT). The spectrum of both mechanical resonators shows two prominent peaks along with the presence of a noise-like continuous spectrum in the background. The frequency gap $\Delta\omega$ between the two peaks is about $2\pi\times40\text{MHz}$ which is close to the value $\tilde{\omega}_+-\tilde{\omega}_-=2J_c$. Therefore, it implies that mechanical energy exchange took place between the two hybridized modes $\hat{\tilde{b}}_\pm$ with their resonance frequency $\tilde{\omega}_\pm$, but the exchange process is highly nonlinear.  Now, for the sake of understanding,  let us assume that the effective optomechanical coupling $\tilde G_\pm$ discussed in the previous section becomes $\tilde G_\pm^{\text{nonlinear}}$ in the strongly nonlinear regime and also assume  $\tilde G_\pm^{\text{nonlinear}}$  becomes non-zero for $\theta\neq n\pi$ and zero for $\theta=n\pi$.  For the case of dynamics plotted in Fig. \ref{fig4}(a), the phonons in the mechanical resonators are being accumulated corresponding to both the modes $\hat{\tilde{b}}_\pm$ since the optical cavity is excited with blue detuned laser and $\theta\neq n\pi$. The sufficiently driven power and the value of the mechanical coupling phase made $\tilde G_+^{\text{nonlinear}}$ comparable to $\tilde G_-^{\text{nonlinear}}$ which creates an ``optical'' path that facilitates the sustained energy exchange process between hybridized modes. On the other hand, the sustained energy exchange process in Fig. \ref{fig3}(a) is suppressed since either $\tilde G_+^{\text{nonlinear}}$ or $\tilde G_-^{\text{nonlinear}}$ dominates, which makes the ``optical'' path weak. This nature of the mechanical dynamics described in Fig. \ref{fig4} impacts the optical intensity dynamics, where it periodically varies from a chaotic state to a regular state and vice versa in the temporal domain, as explained in the next section. 
\begin{figure}[t]
    \centering
    \includegraphics[width=0.7\linewidth]{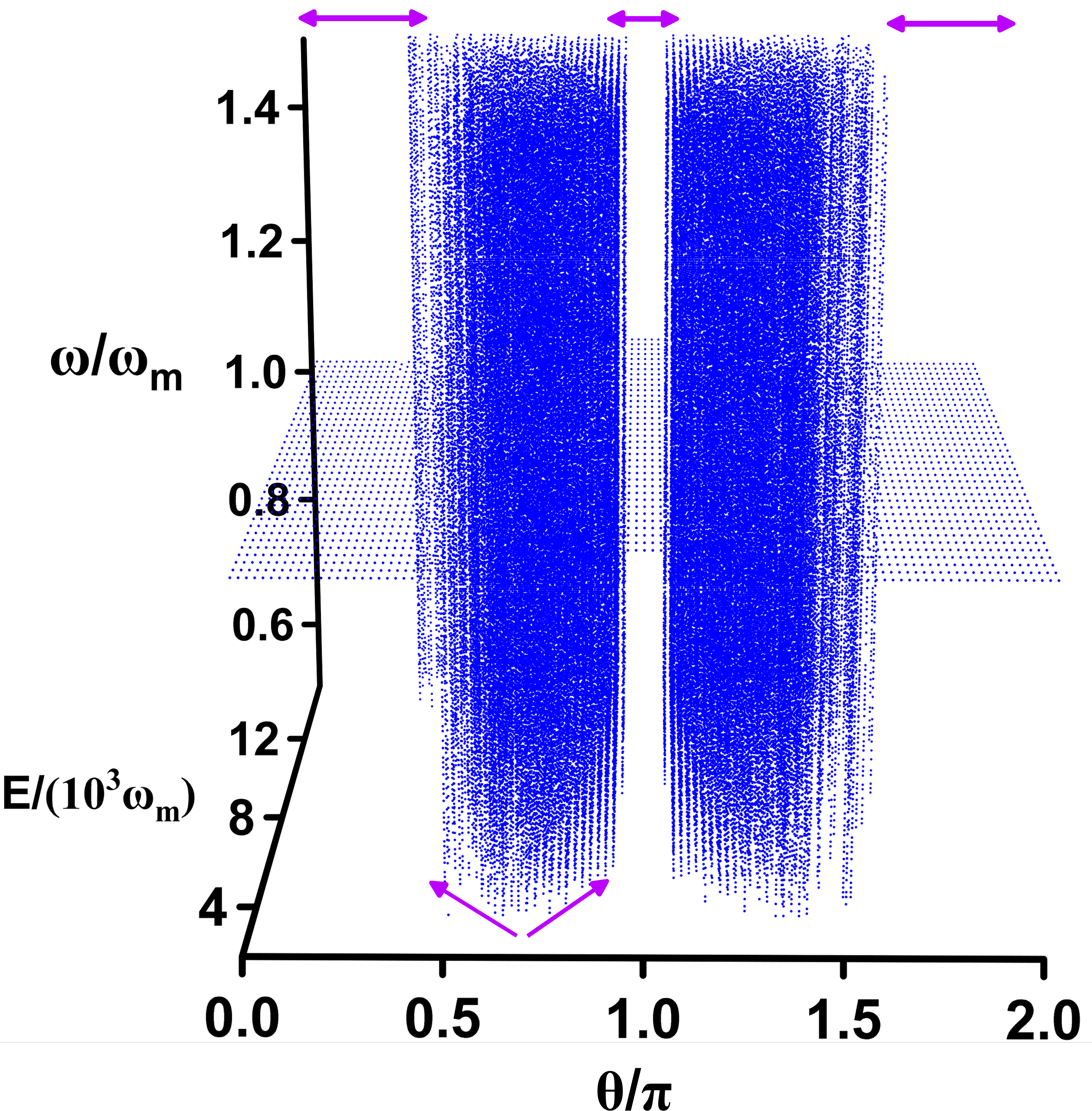}
    \caption{The peaks of the mechanical spectrum $x_1(\omega)\propto\text{Re}
    \;b_{1,2}(\omega)$ under the variation of the driving power $E$ and the phase $\theta$ which provides a comprehensive visualizations of the regime of strong nonlinear dynamics. The horizontal arrows around $\theta=0$, $\theta=\pi$ and $\theta=2\pi$ at the top of the plot shows an approximate region where the irregular nonlinear dynamics is absent and the bottom arrows shows the increase in the range of $\theta$ with increasing $E$.}
    \label{fig6}
\end{figure}
\begin{figure}[t]
    \centering
    \includegraphics[width=0.95\linewidth]{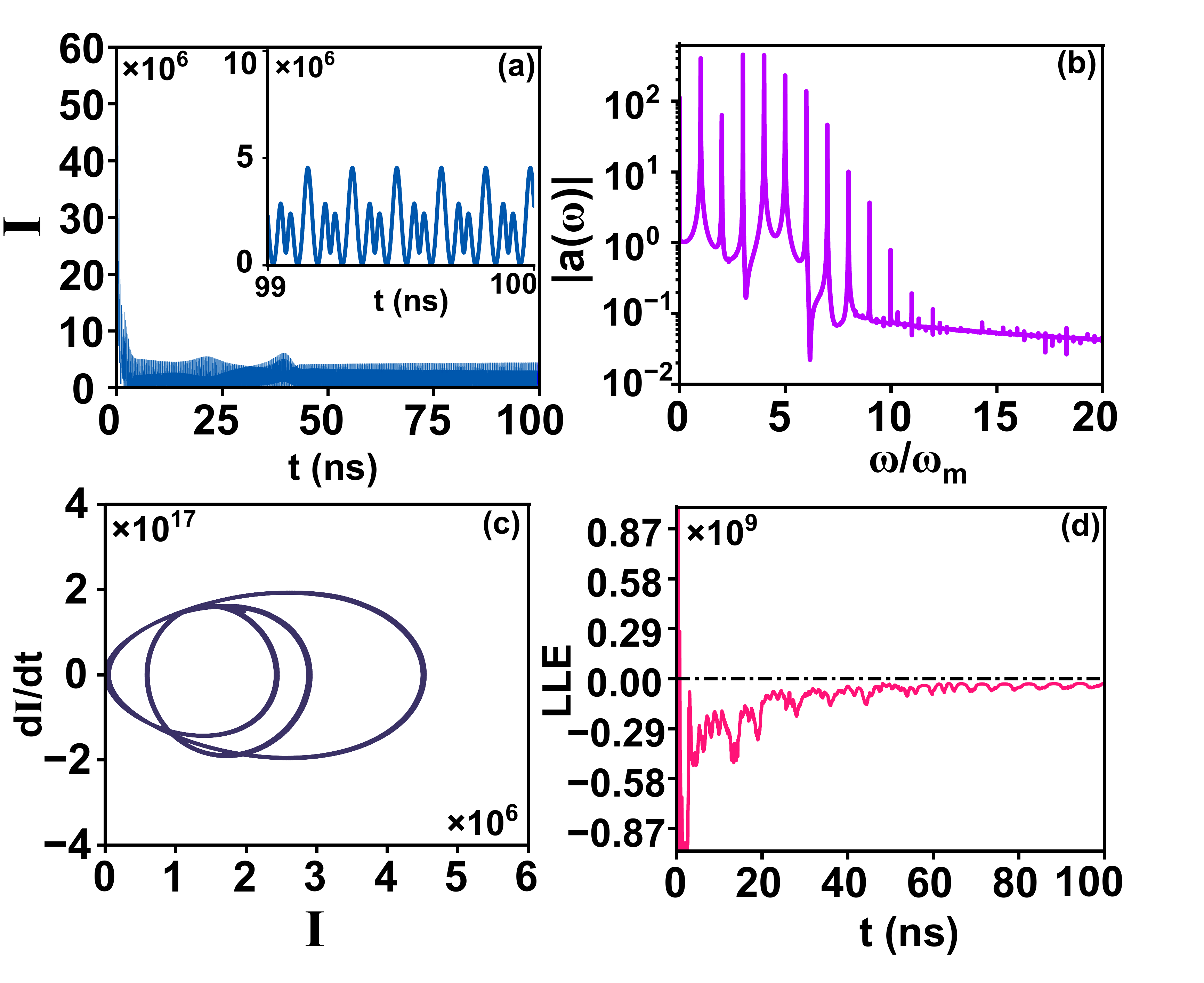}
    \caption{(a) The dynamics of the intracavity intensity $I=|a(t)|^2$ at $E=2000\omega_m$ with $\theta=0.6\pi$ . The inset shows the stable dynamics for a duration of 1 ns. (b) The frequency spectrum $|a(\omega)|$ of the intracavity field consisting of higher-order sidebands. (c) The corresponding optical trajectory in the intensity phase space. (d) The evolution of the LLE with the initial condition of the perturbed quantities are set to $\vec{\delta}(0)=(10^{-12},10^{-12},10^{-12},10^{-12},10^{-12},10^{-12})$.}
    \label{fig7}
\end{figure}
\begin{figure*}
    \centering
    \includegraphics[width=0.9\linewidth]{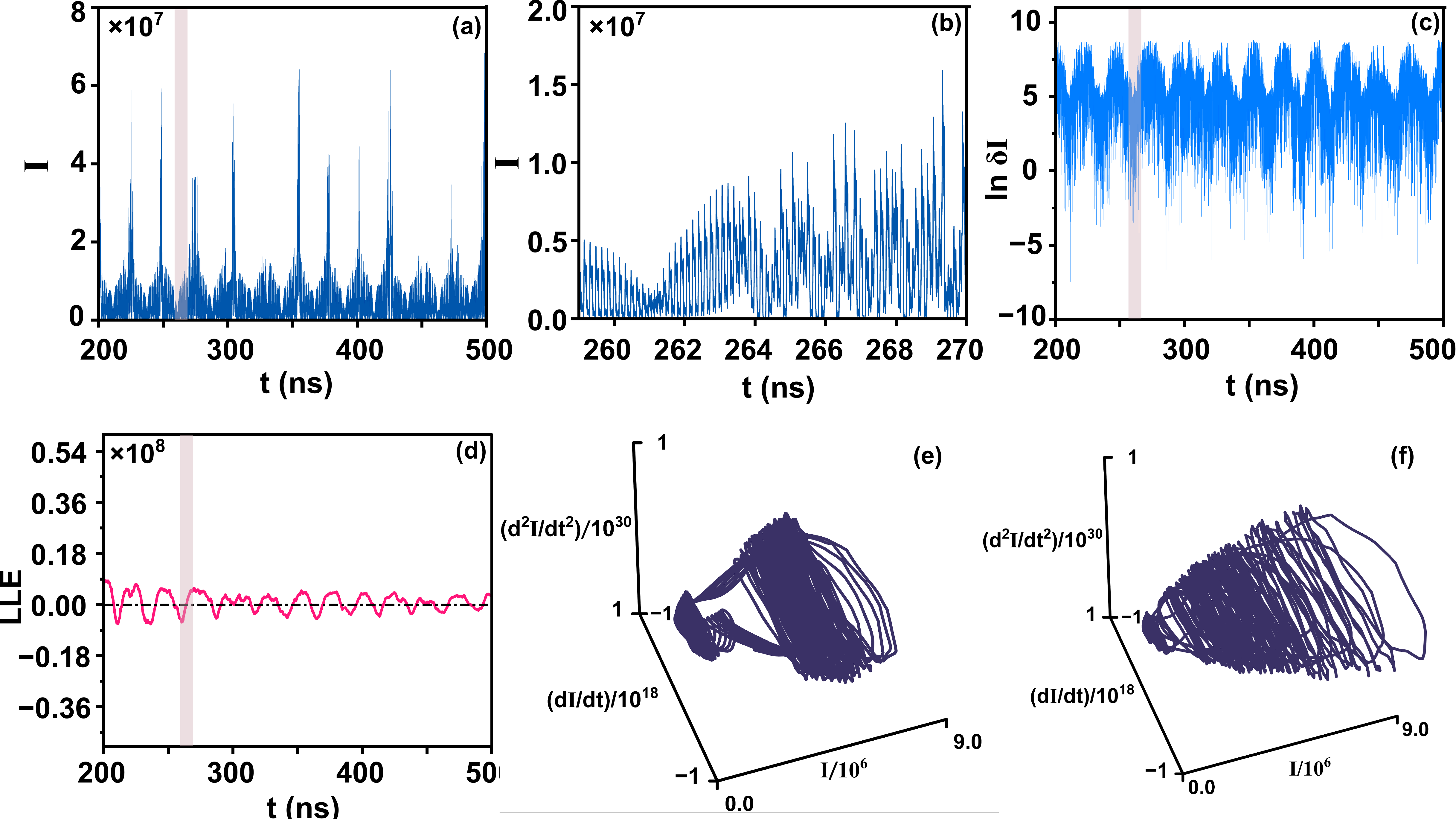}
    \caption{(a) The stable dynamics of the intracavity intensity $I$ at $E=4000\omega_m$ with $\theta=0.6\pi$. (b) The magnified plot indicated by the shaded region in (a) depicts two different nature of the evolution of $I$. (c) The dynamics of ln $\delta I$ over time and (d) the corresponding evolution of the LLE. The optical trajectories in the three dimensional phase space for the case of (e) regular behaviour obtained for time duration $259\text{ns}\to260\text{ns}$ and (f) chaotic behaviour obtained for time duration $264\text{ns}\to265\text{ns}$. The initial condition $\vec{\delta}(0)$ is same as in Fig. \ref{fig7}. }
    \label{fig8}
\end{figure*}

It is convenient to map the frequency response of the mechanical oscillations in the parametric space of the driving power $E$ and the phase $\theta$ to understand better the operational regime of the obtained nonlinear dynamics. Therefore, we plotted the frequency peaks in Fig. \ref{fig6} under the variations of the phase $\theta$ from $0$ to $2\pi$ and the driving power $E$ from $2000\omega_m$ to $15000\omega_m$. The mechanical spectrum is computed after the transient dynamics have died out. The single frequency peaks appear below $E\approx3000\omega_m$ for all values of $\theta$, which says that the strength of the optomechanical nonlinearity is insufficient to observe any irregular dynamics. The equivalent input power $P_{in}$ corresponding to $E=3000\omega_m$ turns out to be around $0.12$W, by assuming $\omega_L=2\pi\times193$ THz. But when the driving power becomes sufficient, a continuous range of frequency peaks indicating strong nonlinearity and erratic evolution of mechanical dynamics appear for a certain range of $\theta$'s, and this range gradually increases as the driving power level becomes higher (follow the bottom arrows in Fig. \ref{fig6}).  The spectrum shows single peaks around $\theta=0,\pi,2\pi$, which are indicated by the horizontal arrows at the top of Fig. \ref{fig6}.  Thus, based on the spectrum in Fig. \ref{fig6}, we observe a strong dependence of the behaviour of mechanical oscillations on the mechanical coupling phase $\theta$. Eventually, the continuous range of frequency peaks would appear for all values of $\theta$' s with a further increase in $E$. 

\section{Dynamics of the intracavity intensity}
In this section, we study the effect of mechanical oscillations on the dynamics of the optical intensity inside the cavity. Particularly, we observe the unique dynamics at three different power levels, from low to high.

\textit{Regular behaviour:} In this scenario, the driving power is set to $E=2000\omega_m$ with fixed phase $\theta=0.6\pi$  and after transient behaviour, the dynamics settle into regular oscillations, which is shown in Fig. \ref{fig7}(a).  Such regular oscillations give rise to higher-order sidebands in the optical spectrum with uniform frequency spacing $\omega_m$ as plotted in Fig. \ref{fig7}(b). The corresponding trajectory of the optical intensity is shown in Fig. \ref{fig7}(c). To verify that the dynamics is regular, we plotted the evolution of the LLE in Fig. \ref{fig7}(d), where we observe the LLE achieves steady negative values.

\textit{Periodic appearances of chaos:} Now, the driving power level is increased to $E=4000\omega_m$, same value as in the previous section and also kept the coupling phase at $\theta=0.6\pi$ such that we can understand the intriguing effects of the mechanical dynamics observed in Fig. \ref{fig4}(a) on the cavity field.  The resultant steady dynamics of the optical intensity inside the cavity at the mentioned operating point is provided in Fig. \ref{fig8}(a), where we see a dip and rise in the intensity value in a periodic fashion. A magnified plot of the dynamics from $259\text{ns}\to270\text{ns}$ is shown in Fig. \ref{fig8}(b), where it shows a time window in which the intensity dynamics evolve in a regular fashion. Then, starting from about $t=264\text{ns}$, the nature of the dynamics changed to a more erratic way, indicating a probable chaotic behaviour.  To quantify this behaviour, we plotted the evolution of the logarithm of $\delta I$ over time in Fig. \ref{fig8}(c), where ln $\delta I$ falls and rises in a periodic manner, which indicates that the perturbed trajectory converges and diverges respectively towards (and from) the original trajectory. The corresponding evolution of LLE is shown in Fig. \ref{fig8}(d), which indicates that the LLE become positive (negative) when the trajectories diverge (converge), which is understandable from all the shaded regions in Fig. \ref{fig8}. The periodicity of the appearance of chaotic and regular dynamics, as calculated from the dynamics, is closer to the frequency gap $\Delta\omega=2J_c$ between the two hybridized modes $\hat{\tilde{b}}_\pm$. This implies that breaking the dark mode through a synthetic magnetic field introduces periodic occurrences of chaotic and regular dynamics, and the mechanical coupling rate plays an essential role in determining the periodicity. The optical intensity trajectories in the three-dimensional phase space also help to visualize the distinct nature between the chaotic and regular dynamics as shown in Fig. \ref{fig8}(e) and \ref{fig8}(f) respectively, where the chaotic dynamics follows a more complex trajectory.  Now, this nature of the dynamics gradually diminishes with the increase of driving power level as full chaotic behaviour starts to dominate because of the increased optomechanical nonlinearity. By observing the dynamics, we roughly estimate the driving power level to be $E=10000\omega_m$, beyond which the full chaotic behaviour governs the dynamics.
\begin{figure}
    \centering
    \includegraphics[width=0.6\linewidth]{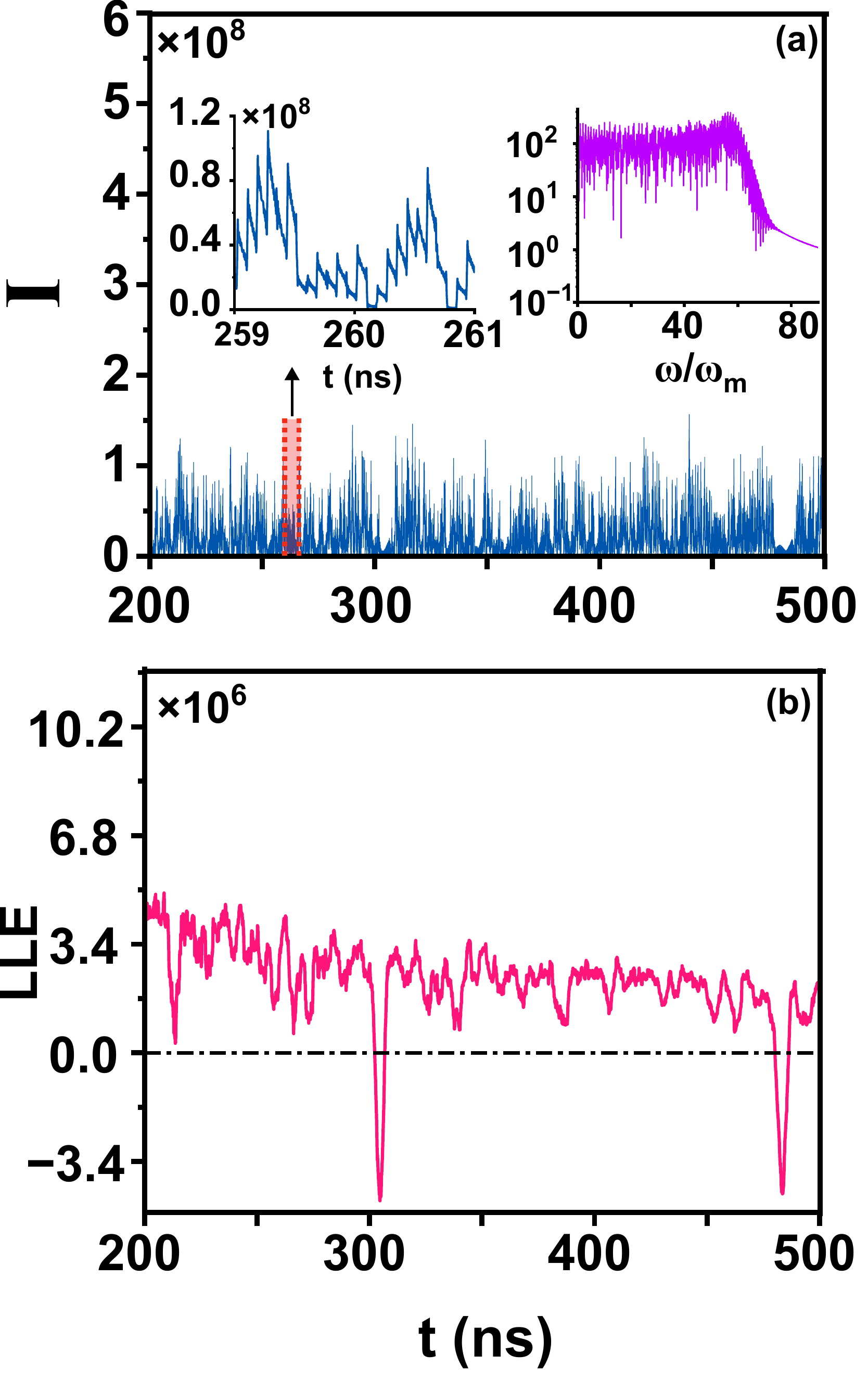}
    \caption{(a) The stable chaotic dynamics of the optical intensity in which the left inset shows the magnified plot of the shaded region and the right inset shows the continuous optical spectrum $|a(\omega)|$. (b) The steady evolution of the positive LLE over time with the initial condition $\vec{\delta}(0)$ kept same as in Fig. \ref{fig7}.}
    \label{fig9}
\end{figure}

\begin{figure}
    \centering
    \includegraphics[width=0.6\linewidth]{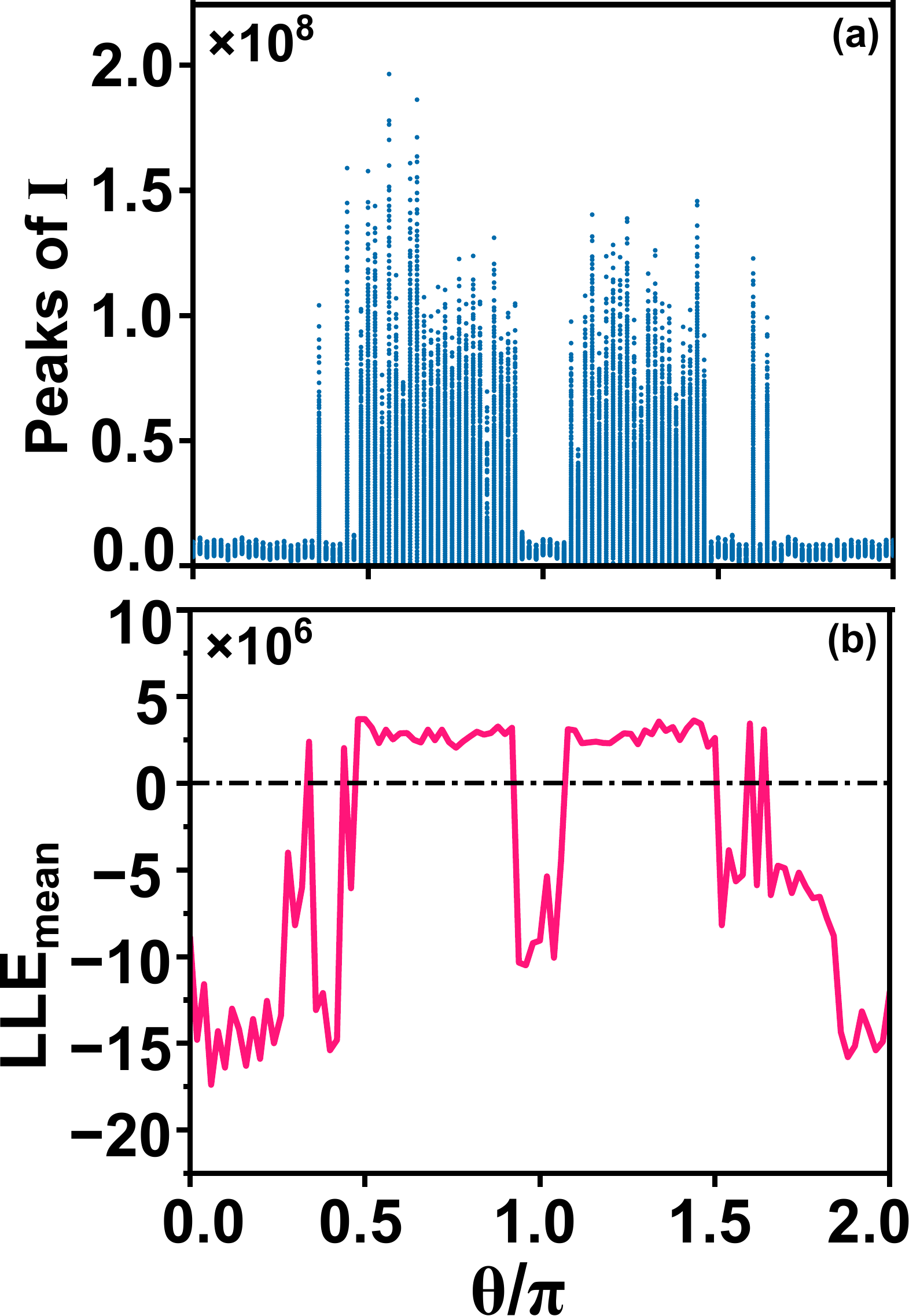}
    \caption{(a) The bifurcation plot of the optical intensity at fixed driving power level $E=12000\omega_m$. (b) The variation of the mean LLE for varying mechanical coupling phase $\theta$.}
    \label{fig10}
\end{figure}

\textit{Chaotic behaviour:} In this case, the driving power level is increased to $E=12000\omega_m$ with the coupling phase fixed at $\theta=0.6\pi$ . The nature of the intensity dynamics in Fig. \ref{fig9}(a) suggests that it is dominated by the chaotic behaviour and the periodic nature of chaos-regular behaviour diminished. The magnified plot of the dynamics in the time window indicated by the shaded region is shown in the left inset of Fig. \ref{fig9} (a). The right inset shows the continuous optical spectrum in this operating regime. The chaotic dynamics are confirmed by the evolution of the LLE over time in Fig. \ref{fig9}(b), where we see that the LLE has positive values except for a few certain time durations where LLE becomes negative. Since the steady evolution has a fluctuating part, one can calculate the mean of the time series to estimate the LLE and the estimation becomes even better if the LLE are calculated for longer maximum simulation time. The chaotic dynamics are being explored for different phase values ranging from $0$ to $2\pi$ through the bifurcation plot in Fig. \ref{fig10}(a). The bifurcation plot is constructed by collecting the peaks of stable intensity dynamics for a time duration of $10\text{ns}$. The relatively heightened portions in the plot indicate the dominant chaotic behaviour and such heightened portions occupy $\theta$ approximately in the range $0.44\pi\lesssim \theta\lesssim0.92\pi$ and $1.08\pi\lesssim \theta\lesssim1.46\pi$ (see Appendix for comparison with that of lower driving power). The mean LLE is obtained for varying $\theta$ in Fig. \ref{fig10}(b), where the heightened portions closely match the positive values of LLE, and the obtained value is roughly three orders less than the mechanical resonance frequency. 

Therefore, in our system, we get a comprehensive idea about the unique nature of the chaotic dynamics of the intracavity intensity when the driving power varies from low to high level and also get an idea of the phase-dependent nature of the dynamics. The chaos appears for the input power of $P_{in}\approx7$ W in a single optomechanical cavity by assuming the same parameter values in our system. On the other hand, the power requirements to induce periodic chaos-regular behaviour in our system is around $0.12$ W, and that of fully chaotic dynamics is around $1.4$ W. Thus, the power requirements to induce chaos have become considerably lower.
\section{Summary}
In summary, we discussed both the nonlinear dynamics of mechanical displacement and the intracavity intensity in the presence of the synthetic magnetic field. The breaking of the dark mode in the presence of a synthetic magnetic field has allowed an irregular and highly nonlinear mechanical energy exchange between the two hybridized mechanical modes, provided the driving optical power is sufficient. The knowledge of the mechanical spectrum is used to map the operational regime of such dynamics under varying driving amplitudes and the mechanical coupling phase. Next, we observed the behaviour of the intensity dynamics inside the cavity, providing an intuitive idea of the effect of strong nonlinear mechanical dynamics on the cavity fields. At a sufficient driving power level, the temporal dynamics is characterized by the periodic appearances of chaotic and regular behaviour in which the mechanical coupling rate governs the periodicity. The evolution of the largest Lyapunov exponent over time is used to quantify this behaviour alongside observing the phase portrait, optical frequency spectrum and the perturbation evolution. Lastly, the dominant chaotic behaviour is observed at a higher driving power level, and its mechanical coupling phase-dependent nature is shown through a bifurcation plot. In terms of practical applicability, the study holds promises in showing low power generation of chaotic signals in an integrated optomechanical system, and the phase of the mechanical coupling provides an additional degree of freedom to control the chaotic behaviour as well.
\appendix
\section{Bifurcation Diagram at $\boldsymbol{E=4000\omega_m}$}
\begin{figure}[t]
    \centering
    \includegraphics[width=0.7\linewidth]{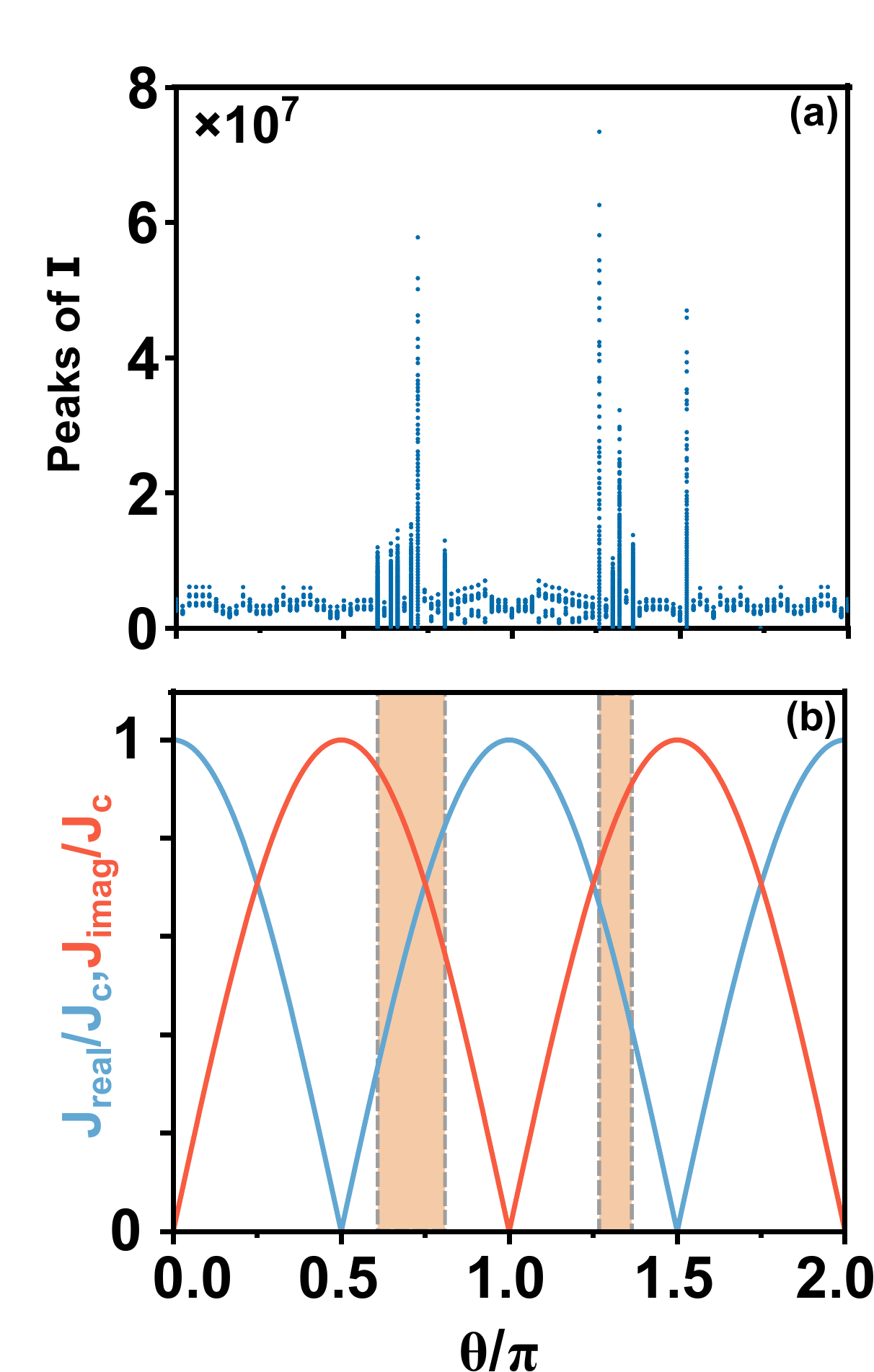}
    \caption{(a) The collected peaks of the optical intensity at driving power level $E=4000\omega_m$ under varying mechanical coupling phase $\theta$. (b) The magnitude of real coupling $J_{\text{real}}$ and imaginary coupling $J_{\text{imag}}$ under varying $\theta$ with the shaded box showing the region of occurrence of the cluster of the heightened lines.}
    \label{fig11}
\end{figure}
Fig. \ref{fig11}(a) shows the bifurcation plot for the low driving power level at $E=4000\omega_m$. Since the optomechanical nonlinearity is weaker than that of the fully chaotic case, it is intuitive to say that the heightened vertical lines would appear sporadically under varying $\theta$ and the mechanical spectrum in Fig. \ref{fig6} also tells the same story. It is also intriguing to describe the chaotic-regular dynamics or fully chaotic dynamics in terms of the strength of the coherent (real) and dissipative (imaginary) nature of mechanical coupling \cite{arwas2022anyonic,cao2024observation} and thereby, a variation of the magnitude of real coupling $J_{\text{real}}=|J_c\text{cos}\theta|$ and imaginary coupling $J_{\text{imag}}=|J_c\text{sin}\theta|$ plotted in Fig. \ref{fig11}(b). The filled box implies the approximate region where the cluster of heightened lines lies. This behaviour conveys that the coexistence of both coherent and dissipative mechanical coupling is essential in determining the dynamics described in Fig \ref{fig4}. Since the occupancy of heightened lines increases with the increase in driving power, the chaotic dynamics persist even if the coupling becomes purely or dominantly dissipative (refer to the bifurcation plot in Fig. \ref{fig10} around $\theta=0.5\pi$ for the fully chaotic case). However, the existence of purely or dominantly coherent coupling does not induce any fully chaotic or periodic regular-chaotic dynamics, as we have seen throughout our study.

\begin{thebibliography}{40}%
\makeatletter
\providecommand \@ifxundefined [1]{%
 \@ifx{#1\undefined}
}%
\providecommand \@ifnum [1]{%
 \ifnum #1\expandafter \@firstoftwo
 \else \expandafter \@secondoftwo
 \fi
}%
\providecommand \@ifx [1]{%
 \ifx #1\expandafter \@firstoftwo
 \else \expandafter \@secondoftwo
 \fi
}%
\providecommand \natexlab [1]{#1}%
\providecommand \enquote  [1]{``#1''}%
\providecommand \bibnamefont  [1]{#1}%
\providecommand \bibfnamefont [1]{#1}%
\providecommand \citenamefont [1]{#1}%
\providecommand \href@noop [0]{\@secondoftwo}%
\providecommand \href [0]{\begingroup \@sanitize@url \@href}%
\providecommand \@href[1]{\@@startlink{#1}\@@href}%
\providecommand \@@href[1]{\endgroup#1\@@endlink}%
\providecommand \@sanitize@url [0]{\catcode `\\12\catcode `\$12\catcode
  `\&12\catcode `\#12\catcode `\^12\catcode `\_12\catcode `\%12\relax}%
\providecommand \@@startlink[1]{}%
\providecommand \@@endlink[0]{}%
\providecommand \url  [0]{\begingroup\@sanitize@url \@url }%
\providecommand \@url [1]{\endgroup\@href {#1}{\urlprefix }}%
\providecommand \urlprefix  [0]{URL }%
\providecommand \Eprint [0]{\href }%
\providecommand \doibase [0]{https://doi.org/}%
\providecommand \selectlanguage [0]{\@gobble}%
\providecommand \bibinfo  [0]{\@secondoftwo}%
\providecommand \bibfield  [0]{\@secondoftwo}%
\providecommand \translation [1]{[#1]}%
\providecommand \BibitemOpen [0]{}%
\providecommand \bibitemStop [0]{}%
\providecommand \bibitemNoStop [0]{.\EOS\space}%
\providecommand \EOS [0]{\spacefactor3000\relax}%
\providecommand \BibitemShut  [1]{\csname bibitem#1\endcsname}%
\let\auto@bib@innerbib\@empty
\bibitem [{\citenamefont {Aspelmeyer}\ \emph {et~al.}(2014)\citenamefont
  {Aspelmeyer}, \citenamefont {Kippenberg},\ and\ \citenamefont
  {Marquardt}}]{aspelmeyer2014cavity}%
  \BibitemOpen
  \bibfield  {author} {\bibinfo {author} {\bibfnamefont {M.}~\bibnamefont
  {Aspelmeyer}}, \bibinfo {author} {\bibfnamefont {T.~J.}\ \bibnamefont
  {Kippenberg}},\ and\ \bibinfo {author} {\bibfnamefont {F.}~\bibnamefont
  {Marquardt}},\ }\bibfield  {title} {\bibinfo {title} {Cavity optomechanics},\
  }\href@noop {} {\bibfield  {journal} {\bibinfo  {journal} {Reviews of Modern
  Physics}\ }\textbf {\bibinfo {volume} {86}},\ \bibinfo {pages} {1391}
  (\bibinfo {year} {2014})}\BibitemShut {NoStop}%
\bibitem [{\citenamefont {Hossein-Zadeh}\ \emph {et~al.}(2006)\citenamefont
  {Hossein-Zadeh}, \citenamefont {Rokhsari}, \citenamefont {Hajimiri},\ and\
  \citenamefont {Vahala}}]{hossein2006characterization}%
  \BibitemOpen
  \bibfield  {author} {\bibinfo {author} {\bibfnamefont {M.}~\bibnamefont
  {Hossein-Zadeh}}, \bibinfo {author} {\bibfnamefont {H.}~\bibnamefont
  {Rokhsari}}, \bibinfo {author} {\bibfnamefont {A.}~\bibnamefont {Hajimiri}},\
  and\ \bibinfo {author} {\bibfnamefont {K.~J.}\ \bibnamefont {Vahala}},\
  }\bibfield  {title} {\bibinfo {title} {Characterization of a
  radiation-pressure-driven micromechanical oscillator},\ }\href@noop {}
  {\bibfield  {journal} {\bibinfo  {journal} {Physical Review A}\ }\textbf
  {\bibinfo {volume} {74}},\ \bibinfo {pages} {023813} (\bibinfo {year}
  {2006})}\BibitemShut {NoStop}%
\bibitem [{\citenamefont {Carmon}\ \emph {et~al.}(2005)\citenamefont {Carmon},
  \citenamefont {Rokhsari}, \citenamefont {Yang}, \citenamefont {Kippenberg},\
  and\ \citenamefont {Vahala}}]{carmon2005temporal}%
  \BibitemOpen
  \bibfield  {author} {\bibinfo {author} {\bibfnamefont {T.}~\bibnamefont
  {Carmon}}, \bibinfo {author} {\bibfnamefont {H.}~\bibnamefont {Rokhsari}},
  \bibinfo {author} {\bibfnamefont {L.}~\bibnamefont {Yang}}, \bibinfo {author}
  {\bibfnamefont {T.~J.}\ \bibnamefont {Kippenberg}},\ and\ \bibinfo {author}
  {\bibfnamefont {K.~J.}\ \bibnamefont {Vahala}},\ }\bibfield  {title}
  {\bibinfo {title} {Temporal behavior of radiation-pressure-induced vibrations
  of an optical microcavity phonon mode},\ }\href@noop {} {\bibfield  {journal}
  {\bibinfo  {journal} {Physical review letters}\ }\textbf {\bibinfo {volume}
  {94}},\ \bibinfo {pages} {223902} (\bibinfo {year} {2005})}\BibitemShut
  {NoStop}%
\bibitem [{\citenamefont {Mondal}\ and\ \citenamefont
  {Debnath}(2023)}]{mondal2023controllable}%
  \BibitemOpen
  \bibfield  {author} {\bibinfo {author} {\bibfnamefont {S.}~\bibnamefont
  {Mondal}}\ and\ \bibinfo {author} {\bibfnamefont {K.}~\bibnamefont
  {Debnath}},\ }\bibfield  {title} {\bibinfo {title} {Controllable
  optical-sideband generation and synchronization in a mechanical gain-loss
  optomechanical system},\ }\href@noop {} {\bibfield  {journal} {\bibinfo
  {journal} {Physical Review A}\ }\textbf {\bibinfo {volume} {108}},\ \bibinfo
  {pages} {023517} (\bibinfo {year} {2023})}\BibitemShut {NoStop}%
\bibitem [{\citenamefont {Carmon}\ \emph {et~al.}(2007)\citenamefont {Carmon},
  \citenamefont {Cross},\ and\ \citenamefont {Vahala}}]{carmon2007chaotic}%
  \BibitemOpen
  \bibfield  {author} {\bibinfo {author} {\bibfnamefont {T.}~\bibnamefont
  {Carmon}}, \bibinfo {author} {\bibfnamefont {M.}~\bibnamefont {Cross}},\ and\
  \bibinfo {author} {\bibfnamefont {K.~J.}\ \bibnamefont {Vahala}},\ }\bibfield
   {title} {\bibinfo {title} {Chaotic quivering of micron-scaled on-chip
  resonators excited by centrifugal optical pressure},\ }\href@noop {}
  {\bibfield  {journal} {\bibinfo  {journal} {Physical review letters}\
  }\textbf {\bibinfo {volume} {98}},\ \bibinfo {pages} {167203} (\bibinfo
  {year} {2007})}\BibitemShut {NoStop}%
\bibitem [{\citenamefont {Navarro-Urrios}\ \emph {et~al.}(2017)\citenamefont
  {Navarro-Urrios}, \citenamefont {Capuj}, \citenamefont {Colombano},
  \citenamefont {Garc{\'\i}a}, \citenamefont {Sledzinska}, \citenamefont
  {Alzina}, \citenamefont {Griol}, \citenamefont {Mart{\'\i}nez},\ and\
  \citenamefont {Sotomayor-Torres}}]{navarro2017nonlinear}%
  \BibitemOpen
  \bibfield  {author} {\bibinfo {author} {\bibfnamefont {D.}~\bibnamefont
  {Navarro-Urrios}}, \bibinfo {author} {\bibfnamefont {N.~E.}\ \bibnamefont
  {Capuj}}, \bibinfo {author} {\bibfnamefont {M.~F.}\ \bibnamefont
  {Colombano}}, \bibinfo {author} {\bibfnamefont {P.~D.}\ \bibnamefont
  {Garc{\'\i}a}}, \bibinfo {author} {\bibfnamefont {M.}~\bibnamefont
  {Sledzinska}}, \bibinfo {author} {\bibfnamefont {F.}~\bibnamefont {Alzina}},
  \bibinfo {author} {\bibfnamefont {A.}~\bibnamefont {Griol}}, \bibinfo
  {author} {\bibfnamefont {A.}~\bibnamefont {Mart{\'\i}nez}},\ and\ \bibinfo
  {author} {\bibfnamefont {C.~M.}\ \bibnamefont {Sotomayor-Torres}},\
  }\bibfield  {title} {\bibinfo {title} {Nonlinear dynamics and chaos in an
  optomechanical beam},\ }\href@noop {} {\bibfield  {journal} {\bibinfo
  {journal} {Nature communications}\ }\textbf {\bibinfo {volume} {8}},\
  \bibinfo {pages} {14965} (\bibinfo {year} {2017})}\BibitemShut {NoStop}%
\bibitem [{\citenamefont {Bakemeier}\ \emph {et~al.}(2015)\citenamefont
  {Bakemeier}, \citenamefont {Alvermann},\ and\ \citenamefont
  {Fehske}}]{bakemeier2015route}%
  \BibitemOpen
  \bibfield  {author} {\bibinfo {author} {\bibfnamefont {L.}~\bibnamefont
  {Bakemeier}}, \bibinfo {author} {\bibfnamefont {A.}~\bibnamefont
  {Alvermann}},\ and\ \bibinfo {author} {\bibfnamefont {H.}~\bibnamefont
  {Fehske}},\ }\bibfield  {title} {\bibinfo {title} {Route to chaos in
  optomechanics},\ }\href@noop {} {\bibfield  {journal} {\bibinfo  {journal}
  {Physical review letters}\ }\textbf {\bibinfo {volume} {114}},\ \bibinfo
  {pages} {013601} (\bibinfo {year} {2015})}\BibitemShut {NoStop}%
\bibitem [{\citenamefont {Thompson}\ \emph {et~al.}(1990)\citenamefont
  {Thompson}, \citenamefont {Stewart},\ and\ \citenamefont
  {Turner}}]{thompson1990nonlinear}%
  \BibitemOpen
  \bibfield  {author} {\bibinfo {author} {\bibfnamefont {J.~M.~T.}\
  \bibnamefont {Thompson}}, \bibinfo {author} {\bibfnamefont {H.~B.}\
  \bibnamefont {Stewart}},\ and\ \bibinfo {author} {\bibfnamefont
  {R.}~\bibnamefont {Turner}},\ }\bibfield  {title} {\bibinfo {title}
  {Nonlinear dynamics and chaos},\ }\href@noop {} {\bibfield  {journal}
  {\bibinfo  {journal} {Computers in Physics}\ }\textbf {\bibinfo {volume}
  {4}},\ \bibinfo {pages} {562} (\bibinfo {year} {1990})}\BibitemShut {NoStop}%
\bibitem [{\citenamefont {Ma}\ \emph {et~al.}(2014)\citenamefont {Ma},
  \citenamefont {You}, \citenamefont {Si}, \citenamefont {Xiong}, \citenamefont
  {Li}, \citenamefont {Yang},\ and\ \citenamefont {Wu}}]{ma2014formation}%
  \BibitemOpen
  \bibfield  {author} {\bibinfo {author} {\bibfnamefont {J.}~\bibnamefont
  {Ma}}, \bibinfo {author} {\bibfnamefont {C.}~\bibnamefont {You}}, \bibinfo
  {author} {\bibfnamefont {L.-G.}\ \bibnamefont {Si}}, \bibinfo {author}
  {\bibfnamefont {H.}~\bibnamefont {Xiong}}, \bibinfo {author} {\bibfnamefont
  {J.}~\bibnamefont {Li}}, \bibinfo {author} {\bibfnamefont {X.}~\bibnamefont
  {Yang}},\ and\ \bibinfo {author} {\bibfnamefont {Y.}~\bibnamefont {Wu}},\
  }\bibfield  {title} {\bibinfo {title} {Formation and manipulation of
  optomechanical chaos via a bichromatic driving},\ }\href@noop {} {\bibfield
  {journal} {\bibinfo  {journal} {Physical Review A}\ }\textbf {\bibinfo
  {volume} {90}},\ \bibinfo {pages} {043839} (\bibinfo {year}
  {2014})}\BibitemShut {NoStop}%
\bibitem [{\citenamefont {Monifi}\ \emph {et~al.}(2016)\citenamefont {Monifi},
  \citenamefont {Zhang}, \citenamefont {{\"O}zdemir}, \citenamefont {Peng},
  \citenamefont {Liu}, \citenamefont {Bo}, \citenamefont {Nori},\ and\
  \citenamefont {Yang}}]{monifi2016optomechanically}%
  \BibitemOpen
  \bibfield  {author} {\bibinfo {author} {\bibfnamefont {F.}~\bibnamefont
  {Monifi}}, \bibinfo {author} {\bibfnamefont {J.}~\bibnamefont {Zhang}},
  \bibinfo {author} {\bibfnamefont {{\c{S}}.~K.}\ \bibnamefont {{\"O}zdemir}},
  \bibinfo {author} {\bibfnamefont {B.}~\bibnamefont {Peng}}, \bibinfo {author}
  {\bibfnamefont {Y.-x.}\ \bibnamefont {Liu}}, \bibinfo {author} {\bibfnamefont
  {F.}~\bibnamefont {Bo}}, \bibinfo {author} {\bibfnamefont {F.}~\bibnamefont
  {Nori}},\ and\ \bibinfo {author} {\bibfnamefont {L.}~\bibnamefont {Yang}},\
  }\bibfield  {title} {\bibinfo {title} {Optomechanically induced stochastic
  resonance and chaos transfer between optical fields},\ }\href@noop {}
  {\bibfield  {journal} {\bibinfo  {journal} {nature photonics}\ }\textbf
  {\bibinfo {volume} {10}},\ \bibinfo {pages} {399} (\bibinfo {year}
  {2016})}\BibitemShut {NoStop}%
\bibitem [{\citenamefont {L{\"u}}\ \emph {et~al.}(2015)\citenamefont {L{\"u}},
  \citenamefont {Jing}, \citenamefont {Ma},\ and\ \citenamefont
  {Wu}}]{lu2015p}%
  \BibitemOpen
  \bibfield  {author} {\bibinfo {author} {\bibfnamefont {X.-Y.}\ \bibnamefont
  {L{\"u}}}, \bibinfo {author} {\bibfnamefont {H.}~\bibnamefont {Jing}},
  \bibinfo {author} {\bibfnamefont {J.-Y.}\ \bibnamefont {Ma}},\ and\ \bibinfo
  {author} {\bibfnamefont {Y.}~\bibnamefont {Wu}},\ }\bibfield  {title}
  {\bibinfo {title} {P t-symmetry-breaking chaos in optomechanics},\
  }\href@noop {} {\bibfield  {journal} {\bibinfo  {journal} {Physical review
  letters}\ }\textbf {\bibinfo {volume} {114}},\ \bibinfo {pages} {253601}
  (\bibinfo {year} {2015})}\BibitemShut {NoStop}%
\bibitem [{\citenamefont {Huang}\ \emph {et~al.}(2021)\citenamefont {Huang},
  \citenamefont {Chen}, \citenamefont {Huang}, \citenamefont {Huang},
  \citenamefont {Liu}, \citenamefont {Chen}, \citenamefont {Luo},\ and\
  \citenamefont {Chen}}]{huang2021tunable}%
  \BibitemOpen
  \bibfield  {author} {\bibinfo {author} {\bibfnamefont {F.}~\bibnamefont
  {Huang}}, \bibinfo {author} {\bibfnamefont {L.}~\bibnamefont {Chen}},
  \bibinfo {author} {\bibfnamefont {L.}~\bibnamefont {Huang}}, \bibinfo
  {author} {\bibfnamefont {J.}~\bibnamefont {Huang}}, \bibinfo {author}
  {\bibfnamefont {G.}~\bibnamefont {Liu}}, \bibinfo {author} {\bibfnamefont
  {Y.}~\bibnamefont {Chen}}, \bibinfo {author} {\bibfnamefont {Y.}~\bibnamefont
  {Luo}},\ and\ \bibinfo {author} {\bibfnamefont {Z.}~\bibnamefont {Chen}},\
  }\bibfield  {title} {\bibinfo {title} {Tunable anti--parity-time-symmetric
  chaos in optomechanics},\ }\href@noop {} {\bibfield  {journal} {\bibinfo
  {journal} {Physical Review A}\ }\textbf {\bibinfo {volume} {104}},\ \bibinfo
  {pages} {L031503} (\bibinfo {year} {2021})}\BibitemShut {NoStop}%
\bibitem [{\citenamefont {Bai}\ \emph {et~al.}(2023)\citenamefont {Bai},
  \citenamefont {Chen}, \citenamefont {Zhang}, \citenamefont {Yan},
  \citenamefont {He}, \citenamefont {Zhang}, \citenamefont {Zhao},\ and\
  \citenamefont {Yu}}]{bai2023tunable}%
  \BibitemOpen
  \bibfield  {author} {\bibinfo {author} {\bibfnamefont {T.-R.}\ \bibnamefont
  {Bai}}, \bibinfo {author} {\bibfnamefont {Z.-D.}\ \bibnamefont {Chen}},
  \bibinfo {author} {\bibfnamefont {J.-Q.}\ \bibnamefont {Zhang}}, \bibinfo
  {author} {\bibfnamefont {D.}~\bibnamefont {Yan}}, \bibinfo {author}
  {\bibfnamefont {Z.-W.}\ \bibnamefont {He}}, \bibinfo {author} {\bibfnamefont
  {S.}~\bibnamefont {Zhang}}, \bibinfo {author} {\bibfnamefont
  {J.}~\bibnamefont {Zhao}},\ and\ \bibinfo {author} {\bibfnamefont
  {Y.}~\bibnamefont {Yu}},\ }\bibfield  {title} {\bibinfo {title} {Tunable
  optical-gain-induced chaotic dynamics in a hidden pt-symmetric optomechanical
  system},\ }\href@noop {} {\bibfield  {journal} {\bibinfo  {journal} {Physical
  Review A}\ }\textbf {\bibinfo {volume} {107}},\ \bibinfo {pages} {033522}
  (\bibinfo {year} {2023})}\BibitemShut {NoStop}%
\bibitem [{\citenamefont {Zhang}\ \emph {et~al.}(2024)\citenamefont {Zhang},
  \citenamefont {Zheng}, \citenamefont {Wang}, \citenamefont {Zhou},\ and\
  \citenamefont {L{\"u}}}]{zhang2024loss}%
  \BibitemOpen
  \bibfield  {author} {\bibinfo {author} {\bibfnamefont {D.-W.}\ \bibnamefont
  {Zhang}}, \bibinfo {author} {\bibfnamefont {L.-L.}\ \bibnamefont {Zheng}},
  \bibinfo {author} {\bibfnamefont {M.}~\bibnamefont {Wang}}, \bibinfo {author}
  {\bibfnamefont {Y.}~\bibnamefont {Zhou}},\ and\ \bibinfo {author}
  {\bibfnamefont {X.-Y.}\ \bibnamefont {L{\"u}}},\ }\bibfield  {title}
  {\bibinfo {title} {Loss-induced chaos in a double-cavity optomechanical
  system},\ }\href@noop {} {\bibfield  {journal} {\bibinfo  {journal} {Physical
  Review A}\ }\textbf {\bibinfo {volume} {109}},\ \bibinfo {pages} {023529}
  (\bibinfo {year} {2024})}\BibitemShut {NoStop}%
\bibitem [{\citenamefont {Zhang}\ \emph {et~al.}(2020)\citenamefont {Zhang},
  \citenamefont {You},\ and\ \citenamefont {L{\"u}}}]{zhang2020intermittent}%
  \BibitemOpen
  \bibfield  {author} {\bibinfo {author} {\bibfnamefont {D.-W.}\ \bibnamefont
  {Zhang}}, \bibinfo {author} {\bibfnamefont {C.}~\bibnamefont {You}},\ and\
  \bibinfo {author} {\bibfnamefont {X.-Y.}\ \bibnamefont {L{\"u}}},\ }\bibfield
   {title} {\bibinfo {title} {Intermittent chaos in cavity optomechanics},\
  }\href@noop {} {\bibfield  {journal} {\bibinfo  {journal} {Physical Review
  A}\ }\textbf {\bibinfo {volume} {101}},\ \bibinfo {pages} {053851} (\bibinfo
  {year} {2020})}\BibitemShut {NoStop}%
\bibitem [{\citenamefont {Zhang}\ \emph {et~al.}(2021)\citenamefont {Zhang},
  \citenamefont {Zheng}, \citenamefont {You}, \citenamefont {Hu}, \citenamefont
  {Wu},\ and\ \citenamefont {L{\"u}}}]{zhang2021nonreciprocal}%
  \BibitemOpen
  \bibfield  {author} {\bibinfo {author} {\bibfnamefont {D.-W.}\ \bibnamefont
  {Zhang}}, \bibinfo {author} {\bibfnamefont {L.-L.}\ \bibnamefont {Zheng}},
  \bibinfo {author} {\bibfnamefont {C.}~\bibnamefont {You}}, \bibinfo {author}
  {\bibfnamefont {C.-S.}\ \bibnamefont {Hu}}, \bibinfo {author} {\bibfnamefont
  {Y.}~\bibnamefont {Wu}},\ and\ \bibinfo {author} {\bibfnamefont {X.-Y.}\
  \bibnamefont {L{\"u}}},\ }\bibfield  {title} {\bibinfo {title} {Nonreciprocal
  chaos in a spinning optomechanical resonator},\ }\href@noop {} {\bibfield
  {journal} {\bibinfo  {journal} {Physical Review A}\ }\textbf {\bibinfo
  {volume} {104}},\ \bibinfo {pages} {033522} (\bibinfo {year}
  {2021})}\BibitemShut {NoStop}%
\bibitem [{\citenamefont {Vanwiggeren}\ and\ \citenamefont
  {Roy}(1998)}]{vanwiggeren1998communication}%
  \BibitemOpen
  \bibfield  {author} {\bibinfo {author} {\bibfnamefont {G.~D.}\ \bibnamefont
  {Vanwiggeren}}\ and\ \bibinfo {author} {\bibfnamefont {R.}~\bibnamefont
  {Roy}},\ }\bibfield  {title} {\bibinfo {title} {Communication with chaotic
  lasers},\ }\href@noop {} {\bibfield  {journal} {\bibinfo  {journal}
  {Science}\ }\textbf {\bibinfo {volume} {279}},\ \bibinfo {pages} {1198}
  (\bibinfo {year} {1998})}\BibitemShut {NoStop}%
\bibitem [{\citenamefont {Cuomo}\ and\ \citenamefont
  {Oppenheim}(1993)}]{cuomo1993circuit}%
  \BibitemOpen
  \bibfield  {author} {\bibinfo {author} {\bibfnamefont {K.~M.}\ \bibnamefont
  {Cuomo}}\ and\ \bibinfo {author} {\bibfnamefont {A.~V.}\ \bibnamefont
  {Oppenheim}},\ }\bibfield  {title} {\bibinfo {title} {Circuit implementation
  of synchronized chaos with applications to communications},\ }\href@noop {}
  {\bibfield  {journal} {\bibinfo  {journal} {Physical review letters}\
  }\textbf {\bibinfo {volume} {71}},\ \bibinfo {pages} {65} (\bibinfo {year}
  {1993})}\BibitemShut {NoStop}%
\bibitem [{\citenamefont {Ren}\ \emph {et~al.}(2013)\citenamefont {Ren},
  \citenamefont {Baptista},\ and\ \citenamefont {Grebogi}}]{ren2013wireless}%
  \BibitemOpen
  \bibfield  {author} {\bibinfo {author} {\bibfnamefont {H.-P.}\ \bibnamefont
  {Ren}}, \bibinfo {author} {\bibfnamefont {M.~S.}\ \bibnamefont {Baptista}},\
  and\ \bibinfo {author} {\bibfnamefont {C.}~\bibnamefont {Grebogi}},\
  }\bibfield  {title} {\bibinfo {title} {Wireless communication with chaos},\
  }\href@noop {} {\bibfield  {journal} {\bibinfo  {journal} {Physical Review
  Letters}\ }\textbf {\bibinfo {volume} {110}},\ \bibinfo {pages} {184101}
  (\bibinfo {year} {2013})}\BibitemShut {NoStop}%
\bibitem [{\citenamefont {Uchida}\ \emph {et~al.}(2008)\citenamefont {Uchida},
  \citenamefont {Amano}, \citenamefont {Inoue}, \citenamefont {Hirano},
  \citenamefont {Naito}, \citenamefont {Someya}, \citenamefont {Oowada},
  \citenamefont {Kurashige}, \citenamefont {Shiki}, \citenamefont {Yoshimori}
  \emph {et~al.}}]{uchida2008fast}%
  \BibitemOpen
  \bibfield  {author} {\bibinfo {author} {\bibfnamefont {A.}~\bibnamefont
  {Uchida}}, \bibinfo {author} {\bibfnamefont {K.}~\bibnamefont {Amano}},
  \bibinfo {author} {\bibfnamefont {M.}~\bibnamefont {Inoue}}, \bibinfo
  {author} {\bibfnamefont {K.}~\bibnamefont {Hirano}}, \bibinfo {author}
  {\bibfnamefont {S.}~\bibnamefont {Naito}}, \bibinfo {author} {\bibfnamefont
  {H.}~\bibnamefont {Someya}}, \bibinfo {author} {\bibfnamefont
  {I.}~\bibnamefont {Oowada}}, \bibinfo {author} {\bibfnamefont
  {T.}~\bibnamefont {Kurashige}}, \bibinfo {author} {\bibfnamefont
  {M.}~\bibnamefont {Shiki}}, \bibinfo {author} {\bibfnamefont
  {S.}~\bibnamefont {Yoshimori}}, \emph {et~al.},\ }\bibfield  {title}
  {\bibinfo {title} {Fast physical random bit generation with chaotic
  semiconductor lasers},\ }\href@noop {} {\bibfield  {journal} {\bibinfo
  {journal} {Nature Photonics}\ }\textbf {\bibinfo {volume} {2}},\ \bibinfo
  {pages} {728} (\bibinfo {year} {2008})}\BibitemShut {NoStop}%
\bibitem [{\citenamefont {Walter}\ and\ \citenamefont
  {Marquardt}(2016)}]{walter2016classical}%
  \BibitemOpen
  \bibfield  {author} {\bibinfo {author} {\bibfnamefont {S.}~\bibnamefont
  {Walter}}\ and\ \bibinfo {author} {\bibfnamefont {F.}~\bibnamefont
  {Marquardt}},\ }\bibfield  {title} {\bibinfo {title} {Classical dynamical
  gauge fields in optomechanics},\ }\href@noop {} {\bibfield  {journal}
  {\bibinfo  {journal} {New Journal of Physics}\ }\textbf {\bibinfo {volume}
  {18}},\ \bibinfo {pages} {113029} (\bibinfo {year} {2016})}\BibitemShut
  {NoStop}%
\bibitem [{\citenamefont {Mathew}\ \emph {et~al.}(2020)\citenamefont {Mathew},
  \citenamefont {Pino},\ and\ \citenamefont {Verhagen}}]{mathew2020synthetic}%
  \BibitemOpen
  \bibfield  {author} {\bibinfo {author} {\bibfnamefont {J.~P.}\ \bibnamefont
  {Mathew}}, \bibinfo {author} {\bibfnamefont {J.~d.}\ \bibnamefont {Pino}},\
  and\ \bibinfo {author} {\bibfnamefont {E.}~\bibnamefont {Verhagen}},\
  }\bibfield  {title} {\bibinfo {title} {Synthetic gauge fields for phonon
  transport in a nano-optomechanical system},\ }\href@noop {} {\bibfield
  {journal} {\bibinfo  {journal} {Nature nanotechnology}\ }\textbf {\bibinfo
  {volume} {15}},\ \bibinfo {pages} {198} (\bibinfo {year} {2020})}\BibitemShut
  {NoStop}%
\bibitem [{\citenamefont {Chen}\ \emph {et~al.}(2021)\citenamefont {Chen},
  \citenamefont {Zhang}, \citenamefont {Shen}, \citenamefont {Zou},
  \citenamefont {Guo},\ and\ \citenamefont {Dong}}]{chen2021synthetic}%
  \BibitemOpen
  \bibfield  {author} {\bibinfo {author} {\bibfnamefont {Y.}~\bibnamefont
  {Chen}}, \bibinfo {author} {\bibfnamefont {Y.-L.}\ \bibnamefont {Zhang}},
  \bibinfo {author} {\bibfnamefont {Z.}~\bibnamefont {Shen}}, \bibinfo {author}
  {\bibfnamefont {C.-L.}\ \bibnamefont {Zou}}, \bibinfo {author} {\bibfnamefont
  {G.-C.}\ \bibnamefont {Guo}},\ and\ \bibinfo {author} {\bibfnamefont {C.-H.}\
  \bibnamefont {Dong}},\ }\bibfield  {title} {\bibinfo {title} {Synthetic gauge
  fields in a single optomechanical resonator},\ }\href@noop {} {\bibfield
  {journal} {\bibinfo  {journal} {Physical review letters}\ }\textbf {\bibinfo
  {volume} {126}},\ \bibinfo {pages} {123603} (\bibinfo {year}
  {2021})}\BibitemShut {NoStop}%
\bibitem [{\citenamefont {Zapletal}\ \emph {et~al.}(2019)\citenamefont
  {Zapletal}, \citenamefont {Walter},\ and\ \citenamefont
  {Marquardt}}]{zapletal2019dynamically}%
  \BibitemOpen
  \bibfield  {author} {\bibinfo {author} {\bibfnamefont {P.}~\bibnamefont
  {Zapletal}}, \bibinfo {author} {\bibfnamefont {S.}~\bibnamefont {Walter}},\
  and\ \bibinfo {author} {\bibfnamefont {F.}~\bibnamefont {Marquardt}},\
  }\bibfield  {title} {\bibinfo {title} {Dynamically generated synthetic
  electric fields for photons},\ }\href@noop {} {\bibfield  {journal} {\bibinfo
   {journal} {Physical Review A}\ }\textbf {\bibinfo {volume} {100}},\ \bibinfo
  {pages} {023804} (\bibinfo {year} {2019})}\BibitemShut {NoStop}%
\bibitem [{\citenamefont {Peano}\ \emph {et~al.}(2015)\citenamefont {Peano},
  \citenamefont {Brendel}, \citenamefont {Schmidt},\ and\ \citenamefont
  {Marquardt}}]{peano2015topological}%
  \BibitemOpen
  \bibfield  {author} {\bibinfo {author} {\bibfnamefont {V.}~\bibnamefont
  {Peano}}, \bibinfo {author} {\bibfnamefont {C.}~\bibnamefont {Brendel}},
  \bibinfo {author} {\bibfnamefont {M.}~\bibnamefont {Schmidt}},\ and\ \bibinfo
  {author} {\bibfnamefont {F.}~\bibnamefont {Marquardt}},\ }\bibfield  {title}
  {\bibinfo {title} {Topological phases of sound and light},\ }\href@noop {}
  {\bibfield  {journal} {\bibinfo  {journal} {Physical Review X}\ }\textbf
  {\bibinfo {volume} {5}},\ \bibinfo {pages} {031011} (\bibinfo {year}
  {2015})}\BibitemShut {NoStop}%
\bibitem [{\citenamefont {Yang}\ \emph {et~al.}(2015)\citenamefont {Yang},
  \citenamefont {Gao}, \citenamefont {Shi}, \citenamefont {Lin}, \citenamefont
  {Gao}, \citenamefont {Chong},\ and\ \citenamefont
  {Zhang}}]{yang2015topological}%
  \BibitemOpen
  \bibfield  {author} {\bibinfo {author} {\bibfnamefont {Z.}~\bibnamefont
  {Yang}}, \bibinfo {author} {\bibfnamefont {F.}~\bibnamefont {Gao}}, \bibinfo
  {author} {\bibfnamefont {X.}~\bibnamefont {Shi}}, \bibinfo {author}
  {\bibfnamefont {X.}~\bibnamefont {Lin}}, \bibinfo {author} {\bibfnamefont
  {Z.}~\bibnamefont {Gao}}, \bibinfo {author} {\bibfnamefont {Y.}~\bibnamefont
  {Chong}},\ and\ \bibinfo {author} {\bibfnamefont {B.}~\bibnamefont {Zhang}},\
  }\bibfield  {title} {\bibinfo {title} {Topological acoustics},\ }\href@noop
  {} {\bibfield  {journal} {\bibinfo  {journal} {Physical review letters}\
  }\textbf {\bibinfo {volume} {114}},\ \bibinfo {pages} {114301} (\bibinfo
  {year} {2015})}\BibitemShut {NoStop}%
\bibitem [{\citenamefont {Fleury}\ \emph {et~al.}(2016)\citenamefont {Fleury},
  \citenamefont {Khanikaev},\ and\ \citenamefont {Alu}}]{fleury2016floquet}%
  \BibitemOpen
  \bibfield  {author} {\bibinfo {author} {\bibfnamefont {R.}~\bibnamefont
  {Fleury}}, \bibinfo {author} {\bibfnamefont {A.~B.}\ \bibnamefont
  {Khanikaev}},\ and\ \bibinfo {author} {\bibfnamefont {A.}~\bibnamefont
  {Alu}},\ }\bibfield  {title} {\bibinfo {title} {Floquet topological
  insulators for sound},\ }\href@noop {} {\bibfield  {journal} {\bibinfo
  {journal} {Nature communications}\ }\textbf {\bibinfo {volume} {7}},\
  \bibinfo {pages} {11744} (\bibinfo {year} {2016})}\BibitemShut {NoStop}%
\bibitem [{\citenamefont {Schmidt}\ \emph {et~al.}(2015)\citenamefont
  {Schmidt}, \citenamefont {Kessler}, \citenamefont {Peano}, \citenamefont
  {Painter},\ and\ \citenamefont {Marquardt}}]{schmidt2015optomechanical}%
  \BibitemOpen
  \bibfield  {author} {\bibinfo {author} {\bibfnamefont {M.}~\bibnamefont
  {Schmidt}}, \bibinfo {author} {\bibfnamefont {S.}~\bibnamefont {Kessler}},
  \bibinfo {author} {\bibfnamefont {V.}~\bibnamefont {Peano}}, \bibinfo
  {author} {\bibfnamefont {O.}~\bibnamefont {Painter}},\ and\ \bibinfo {author}
  {\bibfnamefont {F.}~\bibnamefont {Marquardt}},\ }\bibfield  {title} {\bibinfo
  {title} {Optomechanical creation of magnetic fields for photons on a
  lattice},\ }\href@noop {} {\bibfield  {journal} {\bibinfo  {journal}
  {Optica}\ }\textbf {\bibinfo {volume} {2}},\ \bibinfo {pages} {635} (\bibinfo
  {year} {2015})}\BibitemShut {NoStop}%
\bibitem [{\citenamefont {Lai}\ \emph {et~al.}(2020{\natexlab{a}})\citenamefont
  {Lai}, \citenamefont {Wang}, \citenamefont {Qin}, \citenamefont {Hou},
  \citenamefont {Nori},\ and\ \citenamefont {Liao}}]{lai2020tunable}%
  \BibitemOpen
  \bibfield  {author} {\bibinfo {author} {\bibfnamefont {D.-G.}\ \bibnamefont
  {Lai}}, \bibinfo {author} {\bibfnamefont {X.}~\bibnamefont {Wang}}, \bibinfo
  {author} {\bibfnamefont {W.}~\bibnamefont {Qin}}, \bibinfo {author}
  {\bibfnamefont {B.-P.}\ \bibnamefont {Hou}}, \bibinfo {author} {\bibfnamefont
  {F.}~\bibnamefont {Nori}},\ and\ \bibinfo {author} {\bibfnamefont {J.-Q.}\
  \bibnamefont {Liao}},\ }\bibfield  {title} {\bibinfo {title} {Tunable
  optomechanically induced transparency by controlling the dark-mode effect},\
  }\href@noop {} {\bibfield  {journal} {\bibinfo  {journal} {Physical Review
  A}\ }\textbf {\bibinfo {volume} {102}},\ \bibinfo {pages} {023707} (\bibinfo
  {year} {2020}{\natexlab{a}})}\BibitemShut {NoStop}%
\bibitem [{\citenamefont {Lai}\ \emph {et~al.}(2022)\citenamefont {Lai},
  \citenamefont {Liao}, \citenamefont {Miranowicz},\ and\ \citenamefont
  {Nori}}]{lai2022noise}%
  \BibitemOpen
  \bibfield  {author} {\bibinfo {author} {\bibfnamefont {D.-G.}\ \bibnamefont
  {Lai}}, \bibinfo {author} {\bibfnamefont {J.-Q.}\ \bibnamefont {Liao}},
  \bibinfo {author} {\bibfnamefont {A.}~\bibnamefont {Miranowicz}},\ and\
  \bibinfo {author} {\bibfnamefont {F.}~\bibnamefont {Nori}},\ }\bibfield
  {title} {\bibinfo {title} {Noise-tolerant optomechanical entanglement via
  synthetic magnetism},\ }\href@noop {} {\bibfield  {journal} {\bibinfo
  {journal} {Physical Review Letters}\ }\textbf {\bibinfo {volume} {129}},\
  \bibinfo {pages} {063602} (\bibinfo {year} {2022})}\BibitemShut {NoStop}%
\bibitem [{\citenamefont {Huang}\ \emph {et~al.}(2023)\citenamefont {Huang},
  \citenamefont {Lai},\ and\ \citenamefont {Liao}}]{PhysRevA.108.013516}%
  \BibitemOpen
  \bibfield  {author} {\bibinfo {author} {\bibfnamefont {J.}~\bibnamefont
  {Huang}}, \bibinfo {author} {\bibfnamefont {D.-G.}\ \bibnamefont {Lai}},\
  and\ \bibinfo {author} {\bibfnamefont {J.-Q.}\ \bibnamefont {Liao}},\
  }\bibfield  {title} {\bibinfo {title} {Controllable generation of mechanical
  quadrature squeezing via dark-mode engineering in cavity optomechanics},\
  }\href {https://doi.org/10.1103/PhysRevA.108.013516} {\bibfield  {journal}
  {\bibinfo  {journal} {Phys. Rev. A}\ }\textbf {\bibinfo {volume} {108}},\
  \bibinfo {pages} {013516} (\bibinfo {year} {2023})}\BibitemShut {NoStop}%
\bibitem [{\citenamefont {Tchounda}\ \emph {et~al.}(2023)\citenamefont
  {Tchounda}, \citenamefont {Djorw{\'e}}, \citenamefont {Engo},\ and\
  \citenamefont {Djafari-Rouhani}}]{tchounda2023sensor}%
  \BibitemOpen
  \bibfield  {author} {\bibinfo {author} {\bibfnamefont {S.~M.}\ \bibnamefont
  {Tchounda}}, \bibinfo {author} {\bibfnamefont {P.}~\bibnamefont
  {Djorw{\'e}}}, \bibinfo {author} {\bibfnamefont {S.~N.}\ \bibnamefont
  {Engo}},\ and\ \bibinfo {author} {\bibfnamefont {B.}~\bibnamefont
  {Djafari-Rouhani}},\ }\bibfield  {title} {\bibinfo {title} {Sensor
  sensitivity based on exceptional points engineered via synthetic magnetism},\
  }\href@noop {} {\bibfield  {journal} {\bibinfo  {journal} {Physical Review
  Applied}\ }\textbf {\bibinfo {volume} {19}},\ \bibinfo {pages} {064016}
  (\bibinfo {year} {2023})}\BibitemShut {NoStop}%
\bibitem [{\citenamefont {Fang}\ \emph {et~al.}(2017)\citenamefont {Fang},
  \citenamefont {Luo}, \citenamefont {Metelmann}, \citenamefont {Matheny},
  \citenamefont {Marquardt}, \citenamefont {Clerk},\ and\ \citenamefont
  {Painter}}]{fang2017generalized}%
  \BibitemOpen
  \bibfield  {author} {\bibinfo {author} {\bibfnamefont {K.}~\bibnamefont
  {Fang}}, \bibinfo {author} {\bibfnamefont {J.}~\bibnamefont {Luo}}, \bibinfo
  {author} {\bibfnamefont {A.}~\bibnamefont {Metelmann}}, \bibinfo {author}
  {\bibfnamefont {M.~H.}\ \bibnamefont {Matheny}}, \bibinfo {author}
  {\bibfnamefont {F.}~\bibnamefont {Marquardt}}, \bibinfo {author}
  {\bibfnamefont {A.~A.}\ \bibnamefont {Clerk}},\ and\ \bibinfo {author}
  {\bibfnamefont {O.}~\bibnamefont {Painter}},\ }\bibfield  {title} {\bibinfo
  {title} {Generalized non-reciprocity in an optomechanical circuit via
  synthetic magnetism and reservoir engineering},\ }\href@noop {} {\bibfield
  {journal} {\bibinfo  {journal} {Nature Physics}\ }\textbf {\bibinfo {volume}
  {13}},\ \bibinfo {pages} {465} (\bibinfo {year} {2017})}\BibitemShut
  {NoStop}%
\bibitem [{\citenamefont {Massel}\ \emph {et~al.}(2011)\citenamefont {Massel},
  \citenamefont {Heikkil{\"a}}, \citenamefont {Pirkkalainen}, \citenamefont
  {Cho}, \citenamefont {Saloniemi}, \citenamefont {Hakonen},\ and\
  \citenamefont {Sillanp{\"a}{\"a}}}]{massel2011microwave}%
  \BibitemOpen
  \bibfield  {author} {\bibinfo {author} {\bibfnamefont {F.}~\bibnamefont
  {Massel}}, \bibinfo {author} {\bibfnamefont {T.~T.}\ \bibnamefont
  {Heikkil{\"a}}}, \bibinfo {author} {\bibfnamefont {J.-M.}\ \bibnamefont
  {Pirkkalainen}}, \bibinfo {author} {\bibfnamefont {S.-U.}\ \bibnamefont
  {Cho}}, \bibinfo {author} {\bibfnamefont {H.}~\bibnamefont {Saloniemi}},
  \bibinfo {author} {\bibfnamefont {P.~J.}\ \bibnamefont {Hakonen}},\ and\
  \bibinfo {author} {\bibfnamefont {M.~A.}\ \bibnamefont {Sillanp{\"a}{\"a}}},\
  }\bibfield  {title} {\bibinfo {title} {Microwave amplification with
  nanomechanical resonators},\ }\href@noop {} {\bibfield  {journal} {\bibinfo
  {journal} {Nature}\ }\textbf {\bibinfo {volume} {480}},\ \bibinfo {pages}
  {351} (\bibinfo {year} {2011})}\BibitemShut {NoStop}%
\bibitem [{\citenamefont {Massel}\ \emph {et~al.}(2012)\citenamefont {Massel},
  \citenamefont {Cho}, \citenamefont {Pirkkalainen}, \citenamefont {Hakonen},
  \citenamefont {Heikkil{\"a}},\ and\ \citenamefont
  {Sillanp{\"a}{\"a}}}]{massel2012multimode}%
  \BibitemOpen
  \bibfield  {author} {\bibinfo {author} {\bibfnamefont {F.}~\bibnamefont
  {Massel}}, \bibinfo {author} {\bibfnamefont {S.~U.}\ \bibnamefont {Cho}},
  \bibinfo {author} {\bibfnamefont {J.-M.}\ \bibnamefont {Pirkkalainen}},
  \bibinfo {author} {\bibfnamefont {P.~J.}\ \bibnamefont {Hakonen}}, \bibinfo
  {author} {\bibfnamefont {T.~T.}\ \bibnamefont {Heikkil{\"a}}},\ and\ \bibinfo
  {author} {\bibfnamefont {M.~A.}\ \bibnamefont {Sillanp{\"a}{\"a}}},\
  }\bibfield  {title} {\bibinfo {title} {Multimode circuit optomechanics near
  the quantum limit},\ }\href@noop {} {\bibfield  {journal} {\bibinfo
  {journal} {Nature communications}\ }\textbf {\bibinfo {volume} {3}},\
  \bibinfo {pages} {987} (\bibinfo {year} {2012})}\BibitemShut {NoStop}%
\bibitem [{\citenamefont {Eckmann}\ and\ \citenamefont
  {Ruelle}(1985)}]{eckmann1985ergodic}%
  \BibitemOpen
  \bibfield  {author} {\bibinfo {author} {\bibfnamefont {J.-P.}\ \bibnamefont
  {Eckmann}}\ and\ \bibinfo {author} {\bibfnamefont {D.}~\bibnamefont
  {Ruelle}},\ }\bibfield  {title} {\bibinfo {title} {Ergodic theory of chaos
  and strange attractors},\ }\href@noop {} {\bibfield  {journal} {\bibinfo
  {journal} {Reviews of modern physics}\ }\textbf {\bibinfo {volume} {57}},\
  \bibinfo {pages} {617} (\bibinfo {year} {1985})}\BibitemShut {NoStop}%
\bibitem [{\citenamefont {Araujo}(2019)}]{araujo2019lyapunov}%
  \BibitemOpen
  \bibfield  {author} {\bibinfo {author} {\bibfnamefont {M.~A.}\ \bibnamefont
  {Araujo}},\ }\emph {\bibinfo {title} {Lyapunov exponents and extensivity in
  multiplex networks}},\ \href@noop {} {Ph.D. thesis},\ \bibinfo  {school}
  {University of Aberdeen} (\bibinfo {year} {2019})\BibitemShut {NoStop}%
\bibitem [{\citenamefont {Lai}\ \emph {et~al.}(2020{\natexlab{b}})\citenamefont
  {Lai}, \citenamefont {Huang}, \citenamefont {Yin}, \citenamefont {Hou},
  \citenamefont {Li}, \citenamefont {Vitali}, \citenamefont {Nori},\ and\
  \citenamefont {Liao}}]{lai2020nonreciprocal}%
  \BibitemOpen
  \bibfield  {author} {\bibinfo {author} {\bibfnamefont {D.-G.}\ \bibnamefont
  {Lai}}, \bibinfo {author} {\bibfnamefont {J.-F.}\ \bibnamefont {Huang}},
  \bibinfo {author} {\bibfnamefont {X.-L.}\ \bibnamefont {Yin}}, \bibinfo
  {author} {\bibfnamefont {B.-P.}\ \bibnamefont {Hou}}, \bibinfo {author}
  {\bibfnamefont {W.}~\bibnamefont {Li}}, \bibinfo {author} {\bibfnamefont
  {D.}~\bibnamefont {Vitali}}, \bibinfo {author} {\bibfnamefont
  {F.}~\bibnamefont {Nori}},\ and\ \bibinfo {author} {\bibfnamefont {J.-Q.}\
  \bibnamefont {Liao}},\ }\bibfield  {title} {\bibinfo {title} {Nonreciprocal
  ground-state cooling of multiple mechanical resonators},\ }\href@noop {}
  {\bibfield  {journal} {\bibinfo  {journal} {Physical Review A}\ }\textbf
  {\bibinfo {volume} {102}},\ \bibinfo {pages} {011502} (\bibinfo {year}
  {2020}{\natexlab{b}})}\BibitemShut {NoStop}%
\bibitem [{\citenamefont {Arwas}\ \emph {et~al.}(2022)\citenamefont {Arwas},
  \citenamefont {Gadasi}, \citenamefont {Gershenzon}, \citenamefont {Friesem},
  \citenamefont {Davidson},\ and\ \citenamefont {Raz}}]{arwas2022anyonic}%
  \BibitemOpen
  \bibfield  {author} {\bibinfo {author} {\bibfnamefont {G.}~\bibnamefont
  {Arwas}}, \bibinfo {author} {\bibfnamefont {S.}~\bibnamefont {Gadasi}},
  \bibinfo {author} {\bibfnamefont {I.}~\bibnamefont {Gershenzon}}, \bibinfo
  {author} {\bibfnamefont {A.}~\bibnamefont {Friesem}}, \bibinfo {author}
  {\bibfnamefont {N.}~\bibnamefont {Davidson}},\ and\ \bibinfo {author}
  {\bibfnamefont {O.}~\bibnamefont {Raz}},\ }\bibfield  {title} {\bibinfo
  {title} {Anyonic-parity-time symmetry in complex-coupled lasers},\
  }\href@noop {} {\bibfield  {journal} {\bibinfo  {journal} {Science advances}\
  }\textbf {\bibinfo {volume} {8}},\ \bibinfo {pages} {eabm7454} (\bibinfo
  {year} {2022})}\BibitemShut {NoStop}%
\bibitem [{\citenamefont {Cao}\ \emph {et~al.}(2024)\citenamefont {Cao},
  \citenamefont {Ju}, \citenamefont {Wang}, \citenamefont {Qi}, \citenamefont
  {Liu}, \citenamefont {Peng}, \citenamefont {Chen}, \citenamefont {Zhu},\ and\
  \citenamefont {Li}}]{cao2024observation}%
  \BibitemOpen
  \bibfield  {author} {\bibinfo {author} {\bibfnamefont {P.-C.}\ \bibnamefont
  {Cao}}, \bibinfo {author} {\bibfnamefont {R.}~\bibnamefont {Ju}}, \bibinfo
  {author} {\bibfnamefont {D.}~\bibnamefont {Wang}}, \bibinfo {author}
  {\bibfnamefont {M.}~\bibnamefont {Qi}}, \bibinfo {author} {\bibfnamefont
  {Y.-K.}\ \bibnamefont {Liu}}, \bibinfo {author} {\bibfnamefont {Y.-G.}\
  \bibnamefont {Peng}}, \bibinfo {author} {\bibfnamefont {H.}~\bibnamefont
  {Chen}}, \bibinfo {author} {\bibfnamefont {X.-F.}\ \bibnamefont {Zhu}},\ and\
  \bibinfo {author} {\bibfnamefont {Y.}~\bibnamefont {Li}},\ }\bibfield
  {title} {\bibinfo {title} {Observation of parity-time symmetry in diffusive
  systems},\ }\href@noop {} {\bibfield  {journal} {\bibinfo  {journal} {Science
  Advances}\ }\textbf {\bibinfo {volume} {10}},\ \bibinfo {pages} {eadn1746}
  (\bibinfo {year} {2024})}\BibitemShut {NoStop}%
\end{thebibliography}
%

\end{document}